\documentclass[12pt,letterpaper]{JHEP3} 

 \usepackage{epsfig}

\usepackage{amssymb}

\usepackage{graphicx}
\usepackage{dcolumn}
\usepackage{bm}

\def\l{\left(}
\def\r{\right)}
\def\la{\langle}
\def\ra{\rangle}
\newcommand{\ba}{\begin{array}}
\newcommand{\ea}{\end{array}}
\newcommand{\be}{\begin{equation}}
\newcommand{\ee}{\end{equation}}
\newcommand{\bea}{\begin{eqnarray}}
\newcommand{\eea}{\end{eqnarray}}
\newcommand{\bg}{\begin{gather}}
\newcommand{\eg}{\end{gather}}
\newcommand{\bseq}{\begin{subequations}}
\newcommand{\eseq}{\end{subequations}}

\makeatletter

\def\lsim{\compoundrel<\over\sim}
\def\compoundrel#1\over#2{\mathpalette\compoundreL{{#1}\over{#2}}}
\def\compoundreL#1#2{\compoundREL#1#2}
\def\compoundREL#1#2\over#3{\mathrel
         {\vcenter{\hbox{$\m@th\buildrel{#1#2}\over{#1#3}$}}}}
\makeatother

\title{Annihilation of NMSSM neutralinos in the Sun and neutrino
  telescope limits} 
\author{Sergei Demidov\\
	Institute for Nuclear Research of Russian Academy of
  Sciences,\\ 
  prospect 60-th October 7A, Moscow 117312, Russia.\\
	E-mail: \email{demidov@ms2.inr.ac.ru}}
\author{Olga Suvorova\\
	Institute for Nuclear Research of Russian Academy of Sciences,\\
prospect 60-th October 7A, Moscow 117312, Russia.\\
	E-mail: \email{suvorova@cpc.inr.ac.ru}}

\abstract{
We investigate neutralino dark matter in the framework of NMSSM
performing a scan over its parameter space and calculating neutralino
capture and annihilation rates in the Sun. We discuss the  
prospects of searches for neutralino dark matter in neutrino
experiments depending on neutralino content and its main annihilation
channel. We recalculate the upper limits on neutralino-proton elastic
cross sections directly from neutrino telescopes upper bounds on
annihilation rates in the Sun. This procedure has advantages as
compared with corresponding recalcalations from the limits on muon
flux, namely, it is independent on details of the experiment and the
recalculation coefficients are universal for any kind of WIMP dark
matter models. We derive 90\% c.l. upper limits on neutralino-proton
cross sections from the results of the Baksan Underground Scintillator 
Telescope.  
} 


\begin{document}

\section{Introduction}\label{sec:level1}
Totality of phenomena observed in neighbouring and far cosmos
indicates an existence of unusual forms of matter and energy, which
constitute up to $95\%$ of the energy density of the Universe. 
Nowadays experiments aimed at measurements of relic
anisotropy~\cite{WMAP:06}, gravitation lensing~\cite{GravLin:98} and
dynamics of galaxies and clasters ~\cite{DClaster:01} allow  to
determine quantitatively dark matter density with a high precision and
thereby greatly developed early F.Zwicky hidden mass hypothesis in
the Universe~\cite{Zwicky:33}.   

In more probable cosmological model $\Lambda CDM$ (lambda cold dark
matter) the contribution of baryonic matter in total density is less
than $5\%$, while the amount of dark matter is about 5 times
larger. Experimentally the abundance of dark matter $\Omega_{\chi}$
expressed in units of critical density is fixed in a very narrow range
$0.1109<\Omega_{\chi} h^2<0.1177$~\cite{WMAP:09}, where h is the
Hubble parameter in $100~({\rm km}/{\rm s})/{\rm Mpc}$ units. 

Basic question about nature of dark matter is still open. There are
no candidates in the Standard Model (SM) of particle physics which
could play a role of the collisionless dark matter in the $\Lambda
CDM$ cosmology paradigm. Among the most popular extensions of the SM
are supersymmetry models (SUSY) which rather naturally provide with a
stable WIMP (weakly interacting massive particle~\cite{WIMP:85}) ---
the lightest neutralino. If thermally produced in the early Universe
the lightest neutralino undergoes self-annihilations and freezes out
at some relic density $\Omega_{\chi}$ (for review see, for
example,~\cite{Jungman:96, Silk:05}).

According to the electroweak (EW) theory weak interactions are
mediated by gauge bosons while in SUSY models new particles with mass
of order of EW scale also appear in the intermediate states. Due to
the large mass of the gauge bosons weak interactions have very small
interaction radius and the magnitude of the cross section of weak
processes is a few orders smaller than that of processes mediated by
strong or electromagnetic forces. Correspondingly, annihilation cross
section of WIMP dark matter particles ensures $\Omega_{\chi}$ density
at EW scale. In this scenario WIMP masses are assumed to be between a
few GeV and hundreds of TeV.

In light of claimed spectral peculiarities the latest results of dark
matter searches both in direct detections (DAMA~\cite{DAMA:08}, CDMS 
II~\cite{CDMS:09}) and in cosmic ray experiments of gamma,
electron-positron and proton-antiproton measurements
(WMAP~\cite{WMAP:08}, EGRET \cite{EGRET:06}, FERMI~\cite{FERMI:09},
INTEGRAL~\cite{INTEGRAL:03}, PAMELA~\cite{PAMELA:08},
ATIC~\cite{ATIC:08}) have been interpreted by many authors 
as an evidence of the existence of light dark matter. In particular,
neutralino dark matter in supersymmetric models were recently
discussed in this context in 
papers~\cite{Bottino:2008mf,Bottino:09}. 

In this note we pay attention to neutralinos in a mass range about
    $20-2000$~{${\rm GeV}/c^2$} in the framework of Next to Minimal
    Supersymmetric 
Standard Model (NMSSM). In particular we consider two samples of its
parameter space: mSUGRA-motivated models and general models without
unification of parameters. With application of NMSSMTools
package~\cite{nmssmtools,Ellwanger:2005dv,Ellwanger:2006rn} we perform
a scan over parameter space of the model and calculate capture and
annihilation rates for neutralinos which could be trapped by the Sun. 
We compare the prospects for NMSSM neutralino searches at neutrino
telescopes with that of in direct detection experiments. We also
discuss the NMSSM benchmark points~\cite{Djouadi:2008uw} in context of  
direct searches and data from neutrino telescopes.  Then we consider a
procedure to derive the limits on neutralino-nucleon cross section
from the limits on annihilation rate in spirit of work~\cite{Edsjo:09}
and use the data of the Baksan Underground Scintillator Telescope to
derive the new limits. 

\section{Theoretical frames of the model}\label{sec:level2}

In supersymmetric models the neutralino is a Majorana fermion. The
interactions of such a WIMP with ordinary matter can be both
spin-dependent (SD) and spin-independent (SI). Moreover the
dynamics of the annihilation and scattering processes depends on SUSY
composition of the lightest neutralino, e.g. either it is a 
purely gaugino, higgsino or their mixture. We consider ${\mathbb Z}_3$
symmetric NMSSM model and use
NMSSMTools package~\cite{nmssmtools,Ellwanger:2005dv,Ellwanger:2006rn}
to explore its parameter space for further analysis. 

The relevant part of superpotential $W$ has the form
\bea
W = \lambda\hat{S}\hat{H}_{u}\cdot\hat{H}_{d} +
\frac{1}{3}\kappa\hat{S}^3,
\eea
where $\hat{S}$ is a superfield which is a singlet with respect to SM 
gauge group and $\hat{H}_{u}$ and $\hat{H}_{d}$ are two Higgs
doublets, $\hat{H}_{u} = \l \hat{H}^{+}_{u}, \hat{H}^{0}_{u}\r^T$,
$\hat{H}_{d} = \l \hat{H}^{0}_{d}, \hat{H}^{-}_{d}\r^T$.
Here we denote superfields by letters with hat and their scalar 
components by letters without it. 

Soft SUSY breaking terms for this model read
\begin{eqnarray}
-{\cal L}_{soft} & = & m_{H_{u}}^{2}H^{\dagger}_{u}H_{u} + 
m_{H_{d}}^{2}H^{\dagger}_{d}H_{d} + m_{S}^{2}S^{*}S \\
& + &\l
\lambda A_{\lambda}H_{u}\cdot H_{d}S + \frac{1}{3}\kappa
A_{\kappa}S^{3} + {\rm h.c.}\r 
+ \frac{1}{2}\left[M_{1}\tilde{B}\tilde{B} 
+ M_{2}\tilde{W}^i\tilde{W}^i 
+ M_{3}\tilde{G}^{a}\tilde{G}^{a}\right].
 \nonumber
\end{eqnarray}
Here $m_{H_{u}}$, $m_{H_{d}}$ and $m_{S}$ are soft
scalar masses, $A_{\lambda}$, $A_{\kappa}$ are trilinear coupling
constants 
and $M_{1,2,3}$ are soft gaugino masses. The Higgs sector of the
model
contains the following independent parameters: $\lambda$, $\kappa$,
$A_{\kappa}$, $A_{\lambda}$, $\tan{\beta} = \la H_{u} \ra/\la H_{d}
\ra$ and $\mu\equiv \lambda\la S \ra$. 

In the NMSSM the lightest mass eigenstate of neutralino can be written
as a linear combination
\bea
\chi = Z_{01}\tilde{B} + Z_{02}\tilde{W}^{3} + Z_{03}\tilde{H}_{u}^{0}
+ Z_{04}\tilde{H}_{d}^{0} + Z_{05}\tilde{s}
\eea
of bino $\tilde{B}$, wino $\tilde{W}^{3}$, two higgsinos
$\tilde{H}_{u}^{0}$ and $\tilde{H}_{d}^{0}$ and also singlino
$\tilde{s}$ which is fermion component of the singlet superfield. Here
coefficients $Z_{0i}$ can be chosen as real-valued.

Below we will discuss the prospects of dark matter
searches for two classes of the NMSSM model to be referred as $Set~1$
and $Set~2$. 
\begin{itemize}
\item {\it Set 1}. Here we take mSUGRA-like NMSSM 
with unification of soft SUSY-breaking  parameters which are specified
at GUT scale
\bea
M_{1,2,3} \equiv M_{1/2},\;\;
M_{\tilde{q}_{i}} = M_{H_{i}} \equiv m_{0},\;\;
A_{i} \equiv A_{0},
\eea
where $m_{\tilde{q}_{i}}$ are soft squarks masses and $A_{i}$ are all
  soft trilinear coupling constants except for $A_{\kappa}$. 
We take these parameters in the following ranges: \\
$0~{\rm GeV} <m_{0}< 5000~{\rm GeV}$,
$0~{\rm GeV}<M_{1/2}<4000~{\rm GeV}$, \\
$-4000~{\rm GeV} < A_{0}, A_{\kappa} < 4000~{\rm GeV}$,
${\rm sign}~\mu > 0$, $0.0<\lambda < 0.7$, $1.7 < \tan{\beta} < 54$.\\  
The dark matter in the mSUGRA-like NMSSM  was already 
  discussed in Refs.~\cite{Hugonie:2007vd,Belanger:2008nt}. In this
  case $A_{\kappa}$ is considered as an independent parameter and
  $\kappa$ is obtained from minimization of the scalar potential (see
  Ref.~\cite{Hugonie:2007vd} for details).

\item {\it Set 2}. Here we assume general NMSSM without unification of
  the soft terms and adopt the following ranges for parameters
  specified at SUSY breaking scale:
  $0~{\rm GeV} < \mu, M_1, M_2 < 2000~{\rm GeV}$,\\
  $300~{\rm GeV} < M_3 < 3000~{\rm GeV}$,
  $-2000~{\rm GeV}< A_{\lambda},A_{\kappa} < 2000~{\rm GeV}$, \\
  $0.0<\lambda < 0.7$, $0.0<\kappa < 0.6$, $1.7 < \tan{\beta} < 54$;
  soft sfermion masses and sfermion trilinear terms are varied within
  $\pm 3$~TeV. The ranges of defined parameters allow to construct rather
  light neutralinos. In a view of the prospects for the indirect dark
  matter searches there was related discussion in
  Ref.~\cite{Ferrer:2006hy} for $m_{\chi}<100$~{${\rm GeV}/c^2$},
  while we do not restrict ourselves by only light neutralinos. 
\end{itemize}

We take $m_{t}=171.4$~GeV for the mass of top-quark.
For both model classes we scan over their parameter space by use
NMSSMTools~\cite{nmssmtools,Ellwanger:2005dv,Ellwanger:2006rn}. All 
unphysical or phenomenologically unacceptable models are rejected by
checks for absence of Landau pole up to GUT  scale for $\lambda$ and
$\kappa$, as well for absence of unphysical global minimum of the
scalar potential in Higgs sector and also for absence of colour
breaking minima. Finally, we follow the NMSSMTools implications of the 
experimental constraints (full list of them can be found in
Refs.~~\cite{nmssmtools,Ellwanger:2005dv,Ellwanger:2006rn}). Note,
however, that we do not impose the condition that the SUSY
contribution to the anomalous magnetic moment of muon should explain
the present $3\sigma$ difference between SM prediction and BNL
results. For our further analysis we take the lightest neutralino as
potential dark matter candidates if their relic abundance is
$\Omega_{DM} < 0.3$ and finally, we obtain about $5 \cdot 10^5$
phenomenologically accepted models for each set of parameters
presented above.   

\section{Neutralino-nucleon SI and SD cross sections}\label{sec:level3}
Among relevant physical quantities spin-dependent and spin-independent
neutralino-nucleon scattering cross sections are more probable to be
tested in experimental searches for dark matter interactions. The
expected signal rate has direct dependence on neutralino scattering
cross sections off ordinary nucleons not only in direct detection
experiments but as well as in neutrino telescopes. In the case of 
annihilations of neutralino pairs inside the Sun where solar chemical
composition contains more than $73\%$ of
hydrogen~\cite{Asplund:2009fu}, either SD or SI parts of the cross
section could give dominant contribution to neutralino capture and 
annihilation rates. 

The effective interaction lagrangian responsible for the
spin-dependent part of the neutralino-nuclei cross section is
\bea
{\cal L}_{A} = d_{q}\l\bar{\chi}\gamma^{\mu}\gamma^5\chi\r
\l\bar{q}\gamma_{\mu}\gamma^{5}q\r
\eea
and the expression for SD cross section has the form
\bea
\label{eq:cs_sd}
\sigma_{SD} = \frac{32}{\pi}G_{F}^{2}m_{r}^{2}\Lambda^{2}J(J+1),
\eea
where $m_{r}$ is the reduced neutralino mass $m_{r} =
\frac{m_{\chi}m}{m_{\chi}+m}$, $m$ and $J$ are the mass and the
spin of the nucleus respectively and  
\bea
\Lambda = \frac{1}{J}\l a_{p}\la S_{p}\ra + a_{n}\la S_{n}\ra\r
\eea
where $\la S_{p}\ra$ and $\la S_{n}\ra$ are the spin content of the
proton and neutron while
\bea
a_{p} = \sum_{q}\frac{d_{q}}{\sqrt{2}G_{F}}\Delta_{q}^{(p)},\;\;
a_{n} = \sum_{q}\frac{d_{q}}{\sqrt{2}G_{F}}\Delta_{q}^{(n)}.\;\;
\eea
As default values for the deltas we use
\bea
\Delta_{u}^{(p)} = 0.842,\;\;
\Delta_{d}^{(p)} = -0.427,\;\;
\Delta_{s}^{(p)} = -0.085.
\label{eq:delta}
\eea
The axial-vector coefficients for neutron can be obtained from the
values~(\ref{eq:delta}) by corresponding isospin rotation.

The spin-independent interactions of neutralino with nucleus are
generated by the following effective interaction lagrangian
\bea
{\cal L} = c_{q}\l\bar{\chi}\chi\r\l\bar{q}q\r.
\eea
We do not discuss here the dependence of physical quantities on
parameters of the NMSSM lagrangian and so we do not present explicit 
expressions for $d_{q}$ and $c_{q}$. The details of the calculation of 
these coupling constants from NMSSM lagrangian can be found e.g. in
Ref.~\cite{Belanger:2008sj}. 

The SI cross section of neutralino on nucleus can be written as
\bea
\sigma_{\rm SI} = \frac{4m_{r}^2}{\pi}\l Zf_{p} + (A-Z)f_{n}\r^2,
\eea
where Z and A are the charge and the atomic number of the nuclei,
respectively, and the nucleon formfactors have the form 
\bea
\frac{f_{N}}{m_{N}} = \sum_{q=u,d,s}f_{T_{q}}^{(N)}\frac{c_{q}}{m_{q}}
+ \frac{2}{27}f_{TG}^{(N)}\sum_{q=c,b,t}\frac{c_{q}}{m_{q}},\;\;
{\rm where}\;\;
f_{TG}^{(N)} = 1 - \sum_{q=u,d,s}f_{T_{q}}^{(N)}.
\eea
The parameters $f_{T_{q}}^{(N)}$ are defined from expressions 
\bea
m_{N}f_{T_{q}}^{(N)} = \la N | m_{q}\bar{q}q | N\ra =
m_{q}B_{q}^{(N)}.
\eea
As discussed in several recent papers \cite{Ellis:2008hf,Ellis:2009ka}
the main uncertainty in determination of SI cross section comes from
the pion-nucleon scattering sigma term, $\sigma_{\pi N}$, which
determines the coefficients $f_{T_q}$ as follows
\begin{eqnarray}
f_{T_u}  & = & \frac{2\sigma_{\pi N}}{m_{N}\l 1+ 
\frac{m_{d}}{m_{u}}\r \l 1 + \frac{B_{d}}{B_{u}}\r} ,\;\; 
f_{T_d}  =  \frac{2\sigma_{\pi N}}{m_{N}\l 1+ 
\frac{m_{u}}{m_{d}}\r \l 1 + \frac{B_{u}}{B_{d}}\r},\\
f_{T_s} & = & \frac{\l\frac{m_{s}}{m_{d}}\r\sigma_{\pi N}y}
{m_{N}\l 1+ \frac{m_{u}}{m_{s}}\r} ,\;\; 
y  \equiv  \frac{2B_{s}}{B_{u}+B_{d}}. 
\end{eqnarray}
We use $m_{u}/m_{d}=0.553$ and $m_{s}/m_{d}=18.9$. Following
Ref.~\cite{Belanger:2008sj} we define the quantity
\bea
z=\frac{B_{u}-B_{s}}{B_{d}-B_{s}}\approx 1.49
\eea
and strange quark density
\bea
y = 1-\frac{\sigma_{0}}{\sigma_{\pi N}},
\eea
where 
\bea
\sigma_0 = \frac{m_u+m_d}{2}\l B_u + B_d - 2B_s\r
\eea
and $\sigma_{\pi N}$-term can be written as
\bea
\sigma_{\pi N} = \frac{m_u+m_d}{2}\l B_u + B_d\r.
\eea
The recent estimates show that the value of $\sigma_{\pi N}$ lies in
the interval $55~{\rm MeV}<\sigma_{\pi N} < 73~{\rm MeV}$ and also it
is suggested that $\sigma_0=35\pm 5$~MeV~\cite{sigma_pi}. We use
$\sigma_{\pi N} = 64~{\rm MeV}$ as default value for our calculations.  
We will back to the discussion of influence of known uncertainties in
the nuclear matrix elements on the neutralino annihilation rates in
the next sections. 

\section{Neutralino capture rate in the Sun}\label{sec:level4}
At present the estimated local density of relic dark matter in the
  solar system $\rho^{loc}_{\chi}$ is about $0.3~{\rm 
  GeV}/cm^3$. The speeds of halo WIMPs crossing the ecliptic plane are 
non-relativistic with mean velocity $\bar{v}\sim 270$~km/s (the root-mean-square of the dispersion) 
in the Maxwell-Boltzmann distribution. As far as they are gravitationally
trapped by the Sun, a multiple scattering of WIMP off solar matter is
likely happened with energy losses. Due to these interactions WIMPs
can be captured inside the Sun. Their orbital motion changes to
smaller radius and toward the dense centre, where they finally settle
and accumulate and as well self-annihilate at a distance smaller than
annihilation length. During the lifetime of the Sun these two
processes of capture and annihilation may reach approximate
equilibrium and in this case accumulated number of WIMP does not
change in time. One of the prominent signature of the neutralino 
annihilations is high energy neutrinos generated in decays of
annihilation products, since these neutrino events
can be detected by neutrino telescopes.

Such a scheme has been considered by A.Gould~\cite{Gould:1987ir} and,
{\it e.g.}, later in detailed paper~\cite{Edsjo:2010sun}, see also
references therein. Follow them we can obtain neutralino annihilation
rate from the capture rate in the Sun by solving known evolution
equation for the number of  WIMPs 
\begin{eqnarray}
{{dN} \over {dt}} = {{C_{C}} - {C_{A}}{N^2}},
\label{eq:one}
\end{eqnarray}
where ${C_{C}}$ is the capture rate of WIMP and ${C_{A}}$ is the
specific annihilation rate related to the Sun volume. The second term
in equation (\ref{eq:one}) has the meaning of the rate of WIMPs pair
annihilations: $2\cdot \Gamma_{A} \equiv C_{A} N^2$. The process of
WIMPs evaporation does not included in (\ref{eq:one}) since it is
negligible for WIMP mass larger a few ${\rm
  GeV}/c^2$~~\cite{Gould:1987ir,Griest:1986yu} .

It is easy to show that when equilibrium time $t_{eq}=(C_{C} \cdot
C_{A} )^{-1/2}$ is much less than the live time of the solar system
the annihilation rate is the half of capture rate 
\begin{eqnarray}
\Gamma_{A} = { {C_{C}} \over {2} } \cdot {\rm tanh}^2(t_{\rm
  Sun}/t_{\rm eq}),
\label{eq:two}
\end{eqnarray}
since the number of WIMPs is solved  as
\begin{eqnarray}
N= \sqrt { {C_{\rm C}} \over {C_{\rm A}} } \cdot {\rm tanh}(t_{\rm
  Sun}/t_{\rm eq}).
\end{eqnarray}
The quantity ${C_{\rm A}}$ is determined by
following expression~\cite{Gould:1987ir,Griest:1986yu} 
\begin{eqnarray}
C_{\rm A} = {{\la\sigma_{\rm ann}\cdot v\ra} \over {V_{\rm Sun}}}
\cdot ({ {m_{\chi}} \over {100~{\rm GeV}} })^{3/2}, 
\label{eq:three}
\end{eqnarray}
where $\la\sigma_{ann}\cdot v\ra$ is the neutralino total annihilation 
cross section thermally averaged with the relative neutralino velocity
$v$ and $V_{\rm Sun}=5.7\cdot 10^{27}$~cm$^3$ is the effective volume
of the Sun.  

For calculation of the capture rate we use simplified
model~\cite{Gould:1992} of the 
solar potential which can be parametrized by escape velocity $v_{es}(r)$
on distance $r$ from the center of the Sun as follows
\bea
\label{sun_potential}
v_{es}^2(r) = v_c^2 - \frac{M(r)}{M_{\odot}}(v_c^2-v_s^2), 
\eea
where $v_{c}=1354$~km/s, $v_s=795$~km/s and $M(r)$ is the mass of the
Sun inside a sphere of radius $r$. In this approximation the capture
rate can be written in an analytical
form~\cite{Gould:1987ir,Gould:1992,Jungman:96} 
\bea
C & = & \frac{M_{\odot}\rho_{\chi}\bar{v}}{4\sqrt{6}m_{\chi}^{2}\eta}
\sum_{i}f_{i}\sigma_{i}\frac{(m_{\chi}+m_{i})^{2}}{m_{i}^{2}a} 
\left\{ \frac{2{\rm
    exp}(-a\hat{\eta}^2)}{(1+a)^{1/2}}{\rm erf}(\hat{\eta}) 
- \frac{{\rm
    exp}(-a\hat{\eta}^2)}{(1+a)^{3/2}(A_{c}^{2}-A_s^2)}
    \times\right.\label{c_sun}\\ 
 & &  \left.\times\left\{\l
\hat{A}_{+}\hat{A}_{-}-\frac{1}{2}-\frac{1+a}{a-b}\r \l{\rm
  erf}(\hat{A}_{+}) - {\rm erf}(\hat{A}_{-})\r  + \frac{1}{\sqrt{\pi}} 
\l \hat{A}_{-}{\rm e}^{-\hat{A}_{+}}-\hat{A}_{+}{\rm
  e}^{-\hat{A}_{-}}\r \right\}_{A_{s}}^{A_{c}} \right. \nonumber\\
 & & \left. + \frac{{\rm
    exp}(-b\check{\eta}^2)}{(a-b)(1+b)^{1/2}(A_{c}^{2} - A_{s}^{2})} 
\left\{{\rm e}^{-(a-b)A^2}\left[2{\rm erf}(\check{\eta})-{\rm
      erf}(\check{A}_{+}) + {\rm
      erf}(\check{A}_{-})\right]\right\}_{A_{s}}^{A_{c}}\right\}
\nonumber
\eea
where $m_{i}$ is the mass of $i$-th chemical element, $f_{i}$ is its
mass content in the Sun, $\sigma_{i}$ is total (the sum of SI and SD
contributions) neutralino-nuclei elastic cross section, $A(r) =
\frac{3v_{es}^2(r)}{2\bar{v}^2}\frac{\mu}{\mu_{-}}$, $A_s =
A_{|v_{es}=v_s}$ and $A_c = A_{|v_{es}=v_c}$ and the following 
notations are introduced
\bea
\hat{A}_{\pm} = \hat{A}\pm\hat{\eta},\;\;
\check{A}_{\pm} = \check{A}\pm\check{\eta},\;\;
\hat{A} = A(1+a)^{1/2},\;\; 
\check{A} = A(1+b)^{1/2},\;\; 
a=\frac{m_{\chi}\bar{v}^2}{3E_{0}},\;\;\\
\hat{\eta}=\frac{\eta}{(1+a)^{1/2}},\;\;
\check{\eta}=\frac{\eta}{(1+b)^{1/2}},\;\; 
b=\frac{\mu}{\mu_{+}}a, \;\;
\mu = \frac{m_{\chi}}{m_{i}}, \;\;
\mu_{+} = \frac{\mu+1}{2},
\nonumber
\eea
where
$E_{0} = \frac{3\hbar^2}{2m_{N}R^2}$ and 
$R = \left[0.91\l\frac{m_{i}}{\rm GeV}\r^{1/3} +  0.3\right]\times
10^{-13}$~cm. Here we assume the isothermal spherical model for halo
WIMPs distribution, where $\eta$ is equal to 1.
Expression~(\ref{c_sun}) is obtained with additional assumption the
all chemical elements are distributed 
uniformly in the Sun. This allows to perform analytic integration
over inner part of the Sun and results in the lengthy term in outer
curly brackets. In our calculations we use uniformly distributed solar
    elements obtained from GS98 solar model presented
in~\cite{Asplund:2009fu}. 

Eq.~(\ref{c_sun}) shows explicitly the direct functional dependence of
the capture rate on using approximations and values of scattering 
cross sections, velocity distribution, local dark matter density
and solar model.  In discussion of physical implications of the
calculated capture rate (and, therefore, annihilation rate) one
should bear in mind all related uncertainties. In particularly,
according to recent analysis in Ref.~\cite{Ellis:2009ka}, error in
determination of $\Delta_{s}^{(p)}$ in calculation of SD cross section
results in 10\% uncertanty in annihilation rate, while lack of exact
knowledge of $\sigma_{\pi N}$ can give rather substantial errors (up
to order of magnitude) in annihilation rate if capture rate is
saturated by SI interactions. The accuracy  of using the
approximation~(\ref{sun_potential}) and assumption about uniformly
distributed elements in the Sun with respect to {\it e.g.}  AGSS09
solar profile~\cite{Asplund:2009fu} is within 11\%. Also as it follows
from cosmological N-body simulations~\cite{Nezri:2010} the local dark
matter density $\rho^{loc}_{\chi}$ might be about 20\% larger than
usually adopted value $0.3~{\rm GeV}/{\rm cm}^3$.  

Now we consider the NMSSM benchmark points proposed and discussed in
the paper~\cite{Djouadi:2008uw} in the search strategies at the
LHC. In the Table~\ref{table:NMP} we present the physical quantities
relevant for the dark matter searches in these models: 
\TABLE[!ht]{\footnotesize
\caption{Benchmark NMSSM points.\label{table:NMP}}
\begin{tabular}{|l|c|c|c|c|}
\hline
{\bf Point} & P1,P2 &  P3 & P4 & P5
\\\hline
$m_{\rm WIMP}, {\rm GeV/c^{2}}$  & 208.2 & 208.3 & 90.7 & 70.5
\\\hline
$\Gamma_{A}, {\rm s}^{-1}$  & $3.8\cdot 10^{17}$ & $2.4\cdot 10^{18}$ 
& $1.8\cdot 10^{23}$ & $1.6\cdot 10^{22}$
\\\hline
${\rm annihilation\;channel}$  & $t\bar{t}$ & $t\bar{t}$ & 
$W^{+}W^{-}$ & $b\bar{b}$
\\\hline
$\sigma_{SI,p}, {\rm pb}$  & $2.4\cdot 10^{-10}$ & $6.7\cdot 10^{-10}$
& $6.6\cdot 10^{-7}$ & $1.3\cdot 10^{-7}$
\\\hline
$\sigma_{SD,p}, {\rm pb}$ & $3.4\cdot 10^{-9}$ & $7.4\cdot 10^{-9}$ &
$8.5\cdot 10^{-4}$ & $2.6\cdot 10^{-5}$
\\\hline\hline
\end{tabular}
}
the annihilation rates, the main annihilation channel and also SI and
SD elastic cross sections. We note in passing that the models P1-P3
and P5 correspond to almost pure bino dark matter, while in the models
P5 the lightest neutralino is singlino.  Comparison with the latest
results of the direct searches (see Sec.~\ref{sec:level5}) shows that
even with account of the error bars (see next section for an
illustrative figure) the points P4 and P5 are already
excluded,  while models P1-P3 can not be probed in the nearest future
direct searches and at neutrino experiments.

Annihilation rates $\Gamma_{A}$ calculated in described frames of
{\it Set 1} and {\it Set 2} models are plotted in
Fig.\ref{arate_annih} as a function of neutralino mass. The upper
limits $\Gamma^{UppLim}_A$ obtained at $90\%$ c.l.  
in neutrino experiments of the Baksan~\cite{Baksan:06} and the
AMANDA~\cite{AMANDAII} collaborations are also shown. In
Fig.\ref{arate_annih} we markup points of the NMSSM models  
by colours correspondent to different dominant annihilation channels, 
with more than 0.5 branching ratio in each case. There is obvious
difference between these two SUSY samples from point of view
annihilation branches. Annihilations to $b\overline{b}$ in mSUGRA-like  
model overcomes others for almost all neutralino masses. Consequently,
these $b\overline{b}$-like models can be excluded only by low energy
threshold experiments as the
Baksan~\cite{Baksan:06,Baksan:96,Baksan:97} and  the Super 
Kamiokande~\cite{SuperK:04} telescopes. A role of $\tau^+\tau^-$  
channel is not remarkable in mSUGRA-like model for low neutralino
masses, while we find expected competition of branching ratios of
$b\overline{b}$ and $\tau^{+}\tau^{-}$ for {\it Set 2}. In mass range
near $60-80$~${\rm GeV}/c^2$ there is a decrease in the number of
points due to closeness of possible value of the scalar Higgs mass to
double neutralino mass. However the models inside this mass interval
have a high potential to be seen or excluded by neutrino
telescopes. The same conclusion can be drawn for scalars channels
(i.e., annihilation of neutralinos to scalar and pseudoscalar Higgs
bosons) for {\it Set 2}.  

In the frames of considered models the lightest neutralino could be 
singlino with light mass even less than 50 ${\rm GeV}/c^2$. Not many
singlinos satisfy condition $\Omega_{\chi}<0.3$ and only small amount
of them survive imposing 3$\sigma$ WMAP density cut as it can be seen
in Fig.~\ref{arate_wmap}. However singlinos appear to be among the
most promising candidates in {\it Set 2}  in searching for neutralino 
annihilation signature in neutrino telescopes. To see which
annihilation channel plays dominant role we show the rest of the
models in Fig.~\ref{arate_wmap_annih} separately and respectively to
colours in Fig.~\ref{arate_annih}.  

We note that for neutralino mass less than 80 ${\rm GeV}/c^2$
main annihilation channels are quark and lepton pairs. Neutralino
annihilating mainly to scalars has larger annihilation rate then one
with leptonic channel. As we see the Baksan exclusion line of
$\Gamma^{UppLim}_A$ goes 
yet closely to model candidates. The AMANDA lower (solid)
line of limit could be compared here only with $\tau^{+}\tau^{-}$,
$W^{+}W^{-}$ and $t\bar{t}$ points while for $b\overline{b}$ channels
(green points) their $\Gamma^{UppLim}_A$ limits~\cite{AMANDAII}
(dashed line) are in two orders higher.  

\section{Limits on neutralino-nucleon cross sections}\label{sec:level5}
The basic conclusion followed from the equation~(\ref{eq:one}) is
that annihilation rate $\Gamma_{A}$ is determined by WIMPs scattering
cross sections off solar matter and in equilibrium does not depend on
their annihilation cross sections
\bea
\label{gamma_eq}
\Gamma_{A} = \frac{C_{C}}{2}.
\eea 
In this case annihilation rate~(\ref{eq:two}) can be divided into two
pieces which correspond to different type of contribution (SI or SD)
to neutralino-nuclei interactions
\begin{eqnarray}
{\Gamma_{A}} = {\Gamma_{A}(\sigma_{SI})+\Gamma_{A}(\sigma_{SD})} 
\label{eq:four},
\end{eqnarray}
where $\Gamma_{A}(\sigma_{SI})$ and $\Gamma_{A}(\sigma_{SD})$ are
determined by~(\ref{gamma_eq}) and~(\ref{c_sun}), where one should
keep only SI or SD part of neutralino-nuclei cross section
respectively. To demonstrate the relative importance of SI and SD
contributions we plot their ratio in Fig.~\ref{si_sd_compare}
for {\it Set 1} and {\it Set 2} in dependence on neutralino mass and
dominant annihilation channels (here and below we present in figures
only the models within $3\sigma$ WMAP region for dark matter
abundance). As expected the axial interactions play dominant role for
relatively light neutralinos ($m_{\chi}\lsim 200-500$~GeV) while
annihilation rate for heavier neutralino is saturated in main part by
scalar interactions. Hence for large neutralino masses the
uncertainties in the calculations of the annihilation rate increase. 

As it has been shown by G.~Wikstrom and J.~Edsjo~\cite{Edsjo:09}, if
the WIMP's processes in the Sun reach exact equilibrium one can derive
upper limits on WIMP scattering cross sections from upper limits on
upward going muon fluxes obtained by neutrino telescopes. Consequently
these results simplify comparison with direct WIMP searches via
nuclear recoil reactions.

Early in the Baksan neutralino search~\cite{Baksan:97} it was
mentioned a preference of limits on annihilation rate
$\Gamma_A$ for the purpose of comparison with model prediction because
muon flux $\Phi_{\mu}$ calculated for some model depends on details of
the experiment, {\it e.g.} on its energy threshold. The magnitude of 
annihilation rate gives the scale for neutrino flux from a distant
source (like the Sun at the distance R), 
that is  
\begin{equation}
{{dN_{\nu_j}} \over {dE_{\nu}}}={{\Gamma_A} \over {4 \pi R^2}}
\sum_i B_i{{dN^i_{\nu_j}} \over {dE_{\nu}}}, \qquad 
\nu_j=\nu_{\mu},{\bar \nu}_{\mu},
\label{eq:flxnu}
\end{equation}
where $dN^i_{\nu_j}/dE_{\nu}$ is the differential spectrum of j-th
neutrinos at the surface of the the Sun produced in i-th annihilation
channel. The branching ratios $B_i$ depend on SUSY parameters as we
show above and obviously drive the shape of the final neutrino
spectrum, which could be either mainly "soft"-like from annihilation 
channels into two quarks (e.g. a pair of  $b\overline{b}$) or mainly
"hard"-like from lepton decay channels of two bosons (e.g. a pair of
$W^{+}W^{-}$). 

Definition of an upper limit on neutralino annihilation rate is read 
from the equation~\cite{Baksan:97}
\begin{equation}
  \Gamma^{UppLim}_A = \frac {N^{UppLim}(\vartheta)}{T \cdot
   \varepsilon_{\vartheta}} 
   \times \frac{1}{P_{ann}},
\end{equation}
where $N^{UppLim}(\vartheta)$ is an experimental limit at a given
confidence level on number of events and for a given half-angle of
open cone $\vartheta$ toward the Sun, T is a time of observation, and 
$\varepsilon_{\vartheta}$ is a fraction of total number of neutrino
events collected within the angle $\vartheta$ from expected neutrino
source. A value $P_{ann}$ is neutrino telescope probability to detect
muon per one pair of neutralino annihilation. Here the upward going
muon  is produced in charged current neutrino interactions in the
Earth. To get the probability $P_{ann}$ the detailed Monte Carlo
simulations are required for the detector response with generator of
neutrino energy spectrum~(\ref{eq:flxnu}) and further propagation of
oscillating and interacting neutrinos in matter (see
e.g. Refs.~\cite{Cirelli:2005gh}) and also induced muons on the route
to the detector. From these studies a relation between 
neutralino mass and the angle $\vartheta$ could be found. The latest
results with 90\% c.l. upper limits of annihilation rates in the Sun
are presented by both the AMANDA~\cite{AMANDAII} and the
IceCube~\cite{IceCube} collaborations and as well by the Baksan
Underground Scintillator Telescope in Ref.~\cite{Baksan:06}. 
 
We suggest to implement obtained upper limits on annihilation rate
$\Gamma^{UppLim}_A$ to put upper limits on neutralino SD and SI
scattering cross sections off protons (see also Eq.(15) in
Ref.~\cite{Edsjo:09})
\bea
\sigma^{UppLim}_{SD,p}(m_{\chi})= \lambda^{SD}\l m_{\chi}\r \cdot
\Gamma^{UppLim}_A(m_{\chi})\\ 
\sigma^{UppLim}_{SI,p}(m_{\chi})= \lambda^{SI} \l m_{\chi}\r \cdot
\Gamma^{UppLim}_A(m_{\chi}). 
\eea
Coefficients for these calculations are following
\begin{eqnarray}
\lambda^{SD}\l m_{\chi}\r = {{\sigma_{SD,p}^{\chi}} \over
  {\Gamma_A^{\chi}}(\sigma_{SD})}, \qquad 
\lambda^{SI}\l m_{\chi}\r = {{\sigma_{SI,p}^{\chi}} \over
  {\Gamma_A^{\chi}}(\sigma_{SI})}. 
\label{eq:five}
\end{eqnarray}
They can be obtained from formulae above with the assumption of
dominance either spin-dependent or spin-independent interactions.
Clearly, these coefficients in contrast with ratios
$\Phi_{\mu}/\Gamma_{A}$, $\sigma_{SI,p}/\Phi_{\mu}$ and
$\sigma_{SD,p}/\Phi_{\mu}$ presented in paper~\cite{Edsjo:09}, are
universal in the sense that they do not depend on the experimental
setup (e.g., on energy threshold) and on the model of the WIMP dark
matter, although it still contains uncertainties of astrophysical
values (see Eq.~(\ref{c_sun})). Coefficients $\lambda^{SD}$ and
$\lambda^{SI}$ as functions of neutralino mass $m_{\chi}$ are plotted
in Fig.\ref{s_to_arate}.   

In Table~\ref{table:lambda} we present the upper limits on
neutralino-proton elastic cross sections $\sigma^{UppLim}_{SD,p}$ and 
$\sigma^{UppLim}_{SI,p}$  which we derive from the Baksan 90\%
c.l. upper limits on annihilation rate~\cite{Baksan:06} by use
coefficients~(\ref{eq:five}) which are also shown in
Table~\ref{table:lambda}. The neutralino mass values are the same as
in the Baksan final table~\cite{Baksan:97}. 
\TABLE[!ht]{
\footnotesize
\begin{tabular}{|c|c|c|c|c|c|}
\hline
$m_{\rm WIMP}, {\rm GeV}/c^2$  & $\lambda^{SI}$, pb/s$^{-1}$ &
$\lambda^{SD}$, pb/s$^{-1}$ & $\sigma^{UppLim}_{SI,p}$, pb &
$\sigma^{UppLim}_{SD,p}$, pb 
\\\hline
12.6 & $4.24\cdot 10^{-30}$ & $2.53\cdot 10^{-28}$ & $6.79\cdot
10^{-3}$ & 0.405
\\\hline 
32.6 & $6.95\cdot 10^{-30}$ & $9.96\cdot 10^{-28}$ & $9.73\cdot
10^{-4}$& 0.139
\\\hline 
82.7 & $1.66\cdot 10^{-29}$ & $5.05\cdot 10^{-27}$ & $5.81\cdot
10^{-5}$ & 0.0177
\\\hline 
210.3 & $5.22\cdot 10^{-29}$ & $3.00\cdot 10^{-26}$ & $5.74\cdot
10^{-7}$ & $3.3\cdot 10^{-4}$
\\\hline 
533.9 & $2.13\cdot 10^{-28}$ & $1.87\cdot 10^{-25}$ & $5.33\cdot
10^{-7}$ & $4.68\cdot 10^{-4}$ 
\\\hline 
1356.7 & $1.11\cdot 10^{-27}$ & $1.19\cdot 10^{-24}$ & $1.28\cdot
10^{-6}$ & $1.37\cdot 10^{-3}$ \\\hline 
3462.8 & $6.64\cdot 10^{-27}$ & $7.74\cdot 10^{-24}$ & $3.85\cdot
10^{-6}$ & $4.49\cdot 10^{-3}$
\\\hline\hline
\end{tabular}
\caption{\label{table:lambda}
 Coefficients $\lambda^{SD}$, $\lambda^{SI}$ and the upper limits
 $\sigma^{UppLim}_{SD,p}$, $\sigma^{UppLim}_{SI,p}$ at 90\% c.l. on 
 elastic $\chi$-proton cross sections recalculated from  the Baksan
 90\% c.l. upper limits on annihilation rates.} 
}

Graphical view of the obtained limits on cross sections is shown in
Fig.~\ref{arate_koeff}. Here for comparison (and as well to check our
procedure) we show both the upper limits on SI and SD cross sections
presented by the AMANDA experiment in~\cite{AMANDAII} and our current
recalculations from the AMANDA upper limits on annihilation
rates~\cite{AMANDAII}.  We observe nice coincidence. Therefore
coefficients $\lambda^{SD}$ and $\lambda^{SI}$ presented in
Table~\ref{table:lambda} and in Fig.\ref{s_to_arate} could be used in
recalculation of annihilation rate limits for other neutrino
telescopes (such as the Baikal NT-200~\cite{Baikal:09} and
ANTARES~\cite{ANTARES:09}) in wide range of neutralino masses.

In Figs.~\ref{si_wmap} and~\ref{sd_wmap} neutralino-proton SI and SD
cross sections calculated for NMSSM models are shown with the
strongest experimental upper limits $\sigma^{UppLim}_{SI,p}$ and
$\sigma^{UppLim}_{SD,p}$ obtained at 90\% c.l. in direct detections 
(CDMS~\cite{CDMS:09}, XENON10~\cite{Angle:2007uj},
ZEPLIN-III~\cite{Lebedenko:2008gb} and
PICASSO~\cite{Archambault:2009}) and neutrino telescopes 
(AMANDA~\cite{AMANDAII}, Baksan~\cite{Baksan:06} and
IceCube~\cite{IceCube}). We implement the DMTools~\cite{DMTools:09} 
database to plot listed upper limits. Although main part of the
DAMA~\cite{DAMA:08} allowed ranges are excluded by different direct
experiments, these $3\sigma$ regions with and without ion channeling
are also shown as shadowed. One can see that the SI limits from the
neutrino detectors are able to close the DAMA results for neutralino
mass around $80-100$~GeV/$c^2$. The Baksan upper limits on SD cross
sections close the values covered by the DAMA for neutralino masses
down to 12 GeV/$c^2$. At the lightest mass the Baksan
$\sigma^{UppLim}_{SD,p}$ level is the same as the latest results of
the ZEPLIN-III~\cite{Lebedenko:2008gb} and
PICASSO~\cite{Archambault:2009}. Finally the Baksan limits on SD cross
sections resolve the "gap"~\cite{Edsjo:09} between limits of the
direct detections and predicted SUSY models, here the NMSSM models.

In Figs.~\ref{si_wmap} and~\ref{sd_wmap} we markup the different
neutralino type by colours. Namely we identify the lightest neutralino
as bino if $Z_{01}^{2}>0.5$, as wino if $Z_{02}^{2}>0.5$, as higgsino
if $Z_{03}^2+Z_{04}^2 > 0.5$ and as singlino if $Z_{05}^2 > 0.5$. For
mSUGRA-type models ({\it Set 1}) the lightest neutralino is bino for
almost entire parameter space. However for the {\it Set 2} there are
quite a lot of models in which dark matter can be wino, higgsino or
singlino. Note that in Fig.~\ref{sd_wmap} we drop all models which are
excluded by the limits on SI neutralino-proton cross section shown in
Figs.~\ref{si_wmap} with account of $\sigma_{\pi N}$
uncertainty. Namely we plot only those points which are not excluded
at least for some values of $\sigma_{\pi N}$ from the interval
$55-73$~MeV. Also one can see that population of bino-like neutralinos
in the NMSSM has better prospects to be found or excluded in
experimental search by both direct methods and neutrino
detections. Two benchmark points P4 and P5 are shown with error bars
for SI cross sections. From SI limits in Figs.~\ref{si_wmap} one can
conclude about exclusion of P4 and P5 models by direct detection
experiments. Also the point P4 in Figs.~\ref{sd_wmap} is close to being
excluded by SD limits from the neutrino telescopes. 

\section{Summary and conclusions}\label{sec:level6}
We discussed here the prospects of neutrino telescope searches for
neutralino dark matter in the framework of the NMSSM. In our study we
analyse two types of the NMSSM scenarios, namely, mSUGRA-like models
and general NMSSM without unification of the soft terms and
perform a scan over the NMSSM parameter space with using NMSSMTools
package. We have found that the dominant annihilation channels in the
mSUGRA-like models are very different from those of the general
NMSSM. In case of the mSUGRA models the lightest neutralino is mainly
bino and it annihilates dominantly into $b\bar{b}$ pairs. However,
neutralino annihilating into $W^{+}W^{-}$ and $t\bar{t}$ pairs is also
very promising for searches at neutrino telescopes. For general NMSSM
model we have found expected competition between  $b\bar{b}$ and
$\tau^{+}\tau^{-}$ channels of annihilation for neutralino mass less
than mass of W-boson, while for heavier neutralino $W^{+}W^{-}$ and
$t\bar{t}$ channels are comparable with those into scalar pairs. We
found that NMSSM neutralinos with mass less about $200~{\rm GeV}/c^2$
scatter off nucleons in the Sun mainly by axial interactions and with
increasing of the mass value the part of the scalar interactions also
increases such that at masses higher than 500~$\rm GeV/c^2$ it is
almost spin-independent scattering.  In general, for both types of
NMSSM models singlino and bino are the most probable models to be
tested either at the neutrino telescopes with low energy threshold
through neutralino axial-vector (SD) interactions in the Sun or in the
direct detection experiments sensible to scalar (SI) neutralino
interactions. 

Depending on neutralino properties there are essentially different
prospects in search for them at neutrino telescopes with low and high
energy thresholds. The type of annihilation products determines
fractions of low (soft) and high (hard) energy spectra of outcoming
particles including neutrinos. Probability of neutrino telescope to
detect neutrinos from predicted neutralino models absorbs dependence
on energy threshold of the detector for neutrino induced upward going 
muons. Presently among the operating neutrino telescopes the Baksan
Underground Scintillator Telescope and the Super Kamiokande have low
energy thresholds (about 1 $\rm GeV$). The Baksan 90\% c.l. upper
limits on annihilation rate in the Sun result from absence neutralino
signal in all expected mass range from dozen ${\rm GeV}/c^2$. The
large detectors with high thresholds e.g. AMANDA or IceCube present
the upper limits on annihilation rates for separated annihilation
branches with soft and hard neutrino spectra. Therefore all limits can
be compared only in context of a particular neutralino annihilation
channel. 

Finally, we have discussed a relation between the limits on neutralino
annihilation rate in the Sun and the limits on neutralino-proton SI
and SD cross section. Under assumption of equilibrium of the
neutralino processes in the Sun we derive 90\% c.l. upper limits on SI 
and SD neutralino-proton cross sections from the upper limits on
annihilation rates for the Baksan Underground Scintillator Telescope. 
Also we tabulate the relevant coefficients for calculation the above
limits, since they are independent on the details of a particular
experiment and the type of WIMP dark matter model. 

\begin{acknowledgments}
We thank S.P.~Mikheev for useful remarks and discussions.
The work was supported in part by the Russian Found for Basic Research  
Grant 09-02-00163a. The work of S.D. was supported by the Russian
Found for Basic Research Grant 08-02-00768a, grants NS-5525.2010.2,
MK-4317.2009.2, by the Federal Agency for Science and Innovation under
state contract 02.740.11.0244, by the Federal Agency for Education
under state contracts P520 and P2598, by the Russian Science Support
Foundation. Numerical part of the work was performed on the
Computational cluster of the Theoretical Division of INR RAS. 
\end{acknowledgments}

%
\FIGURE[htb]{
\includegraphics[width=0.9\columnwidth,height=0.6\columnwidth]{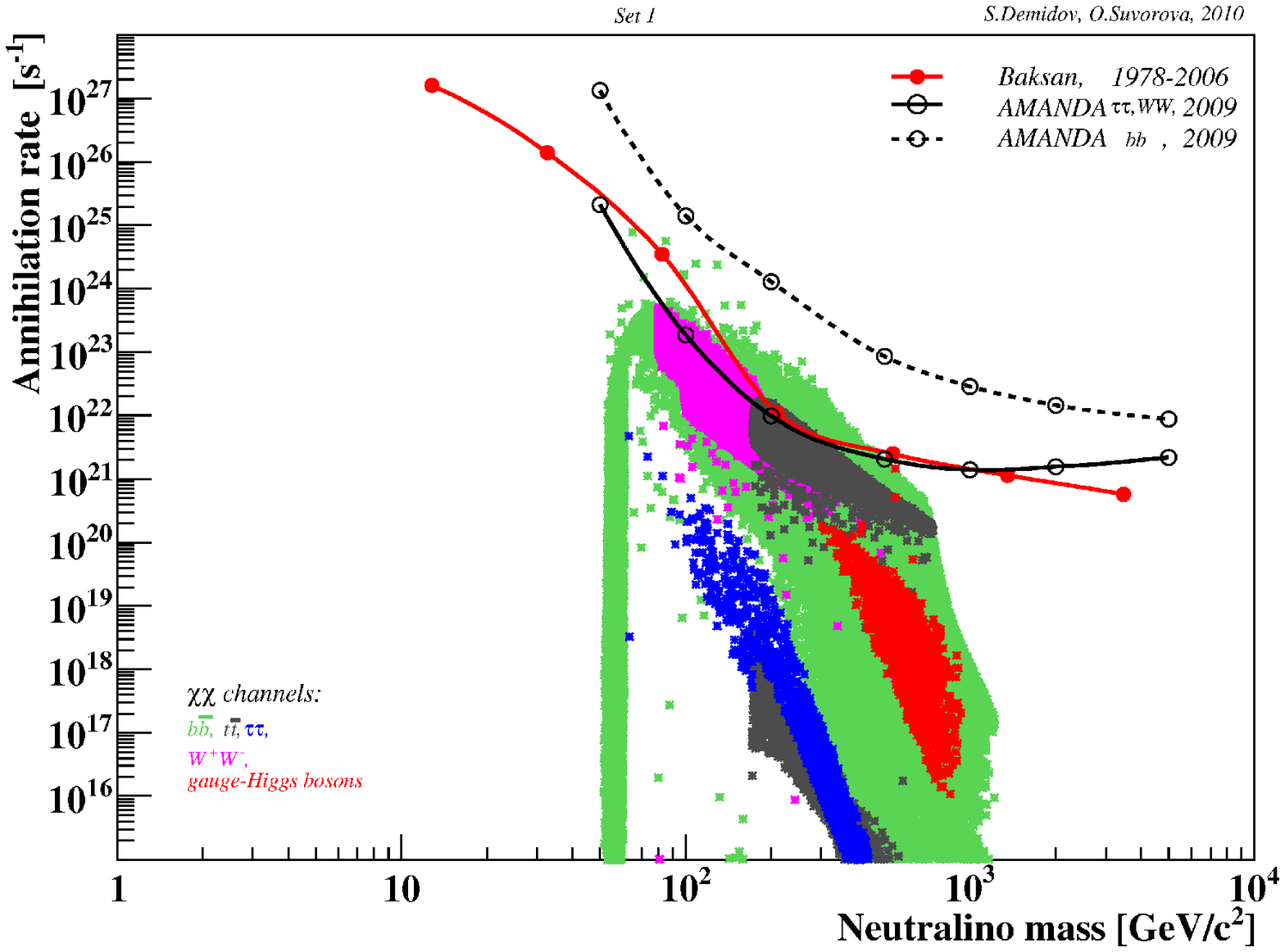} 
\includegraphics[width=0.9\columnwidth,height=0.6\columnwidth]{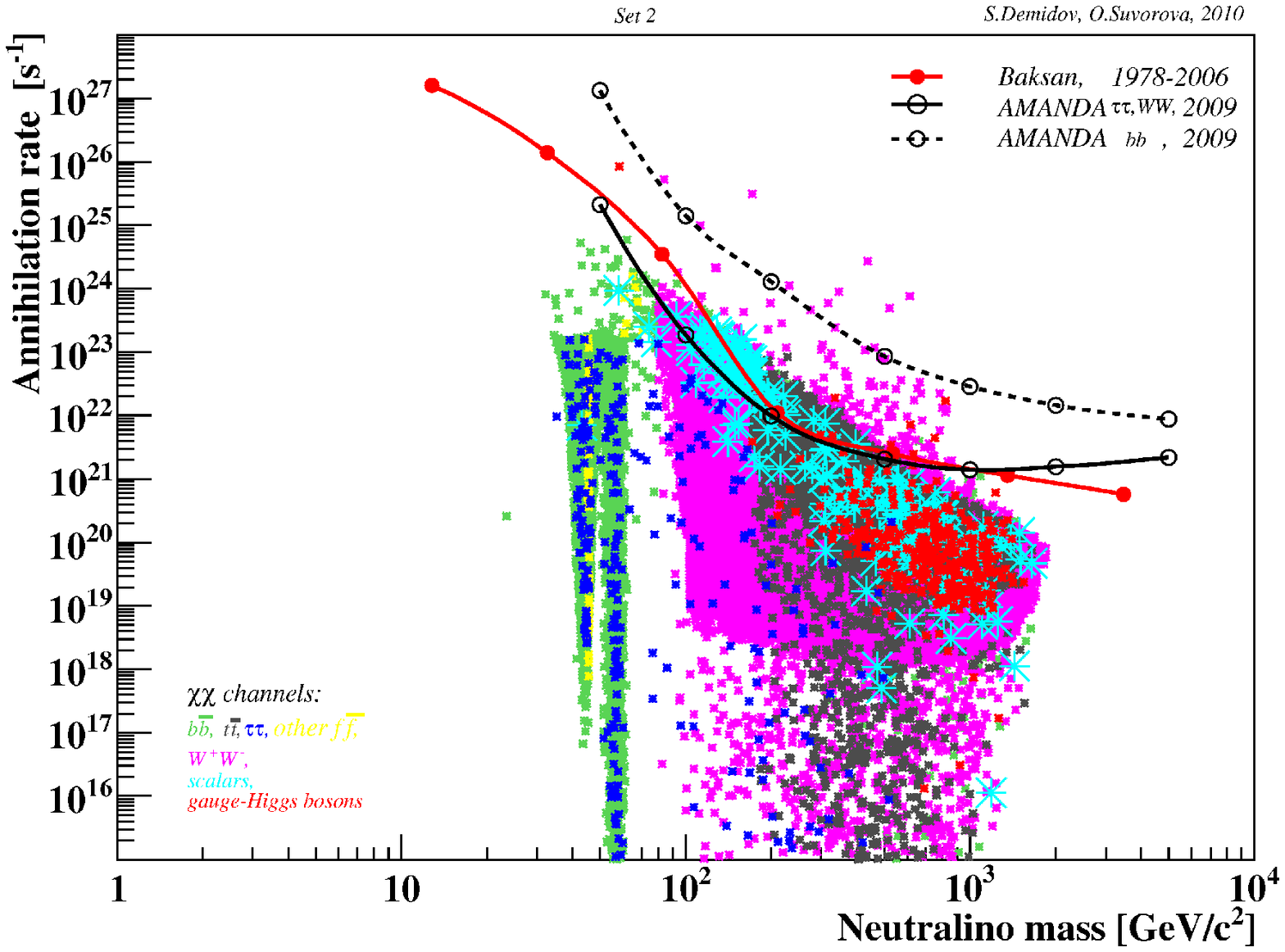}
\caption{\label{arate_annih}
Annihilation rates vs neutralino mass for {\it Set 1} (up) and {\it
  Set 2} (bottom) of parameters. Here all points are in the interval
$0<\Omega_{\chi} < 0.3$. We show the dominant annihilation channel
(${\rm Br} > 0.5$) by different colours. 
}
}
\FIGURE[htb]{
\includegraphics[width=0.9\columnwidth,height=0.6\columnwidth]{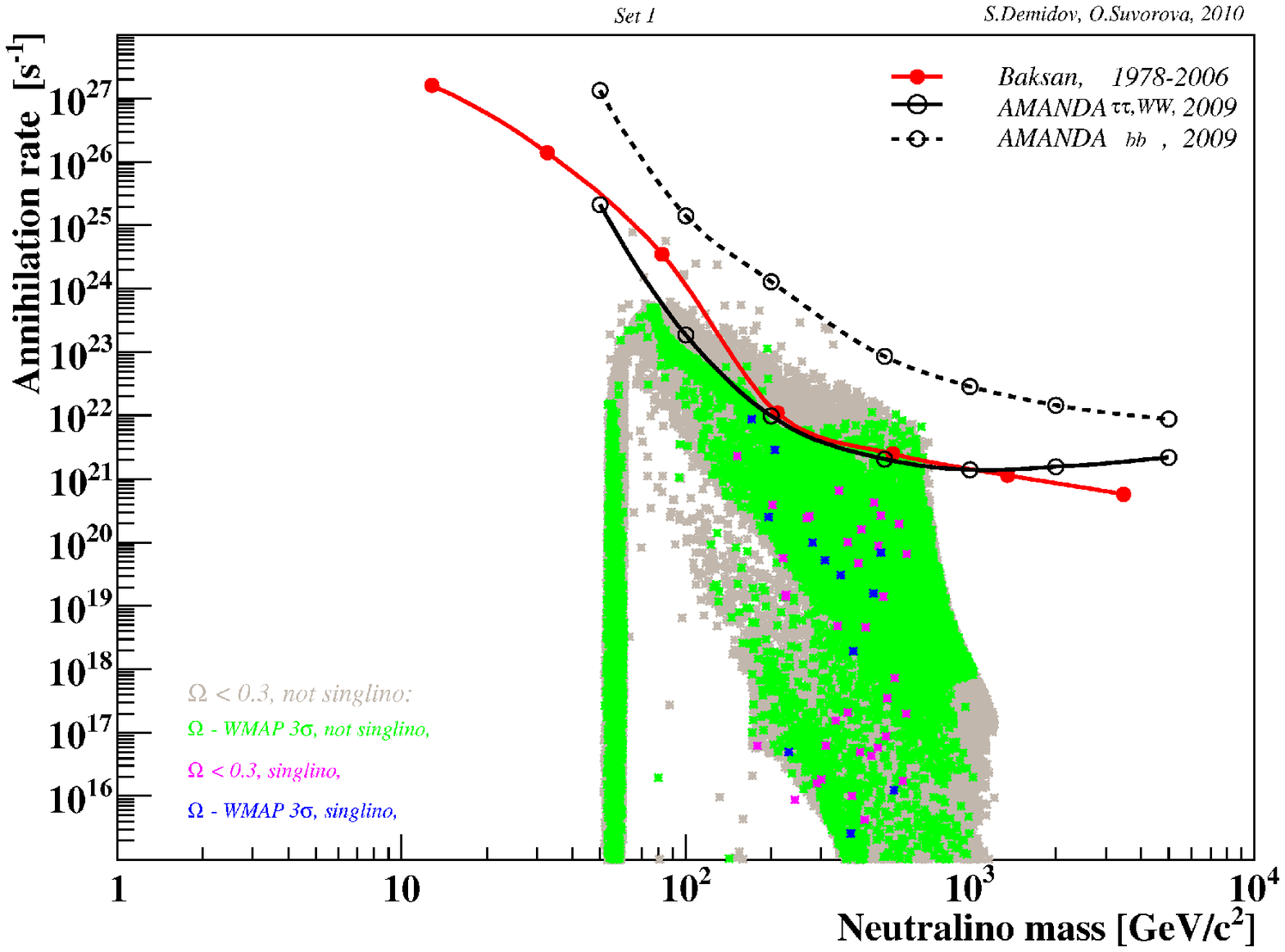}
\includegraphics[width=0.9\columnwidth,height=0.6\columnwidth]{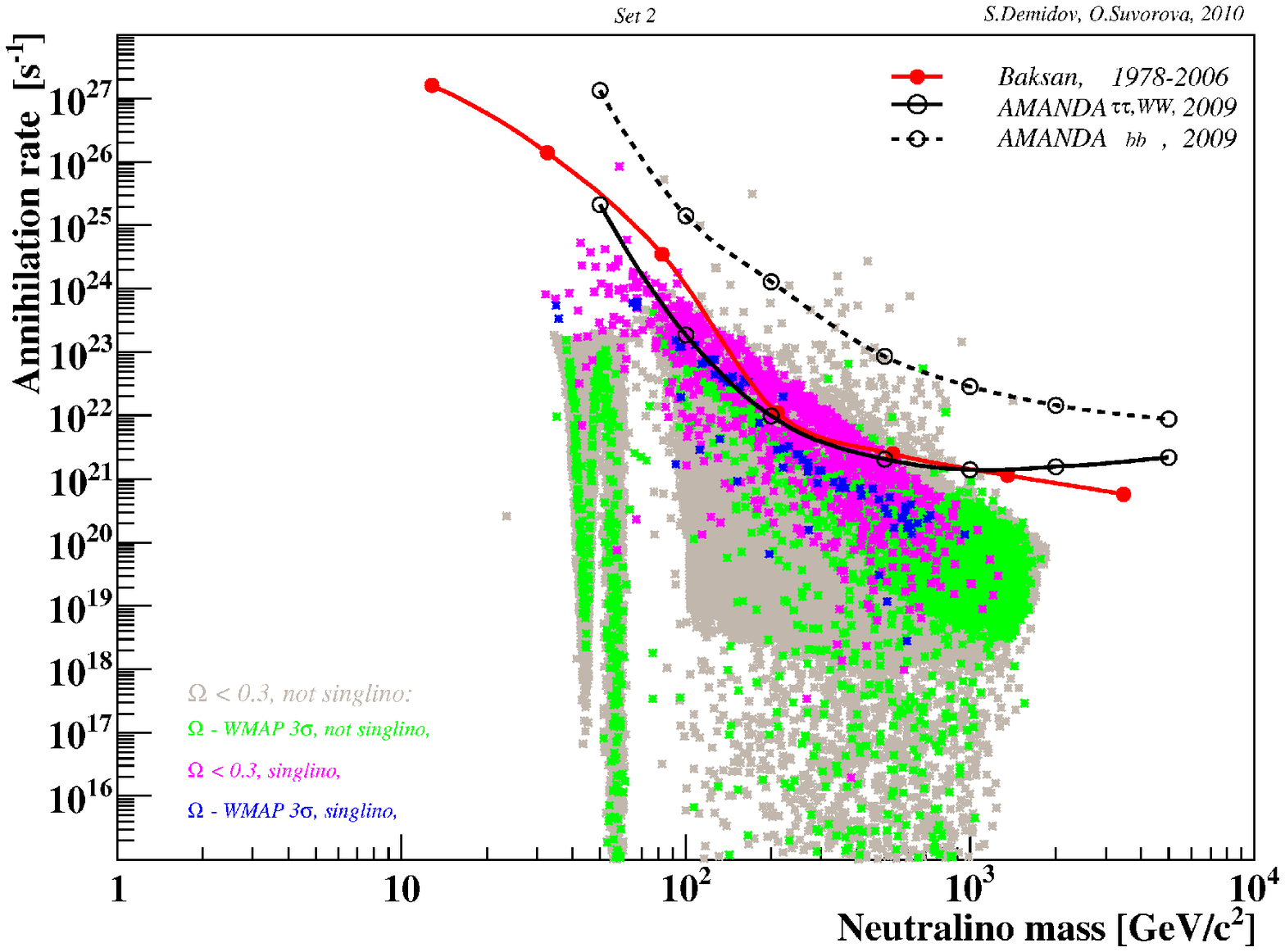}
\caption{\label{arate_wmap}
Annihilation rates vs neutralino mass for {\it Set 1} (up) and {\it
  Set 2} (bottom) of parameters. }
}
\FIGURE[htb]{
\includegraphics[width=0.9\columnwidth,height=0.6\columnwidth]{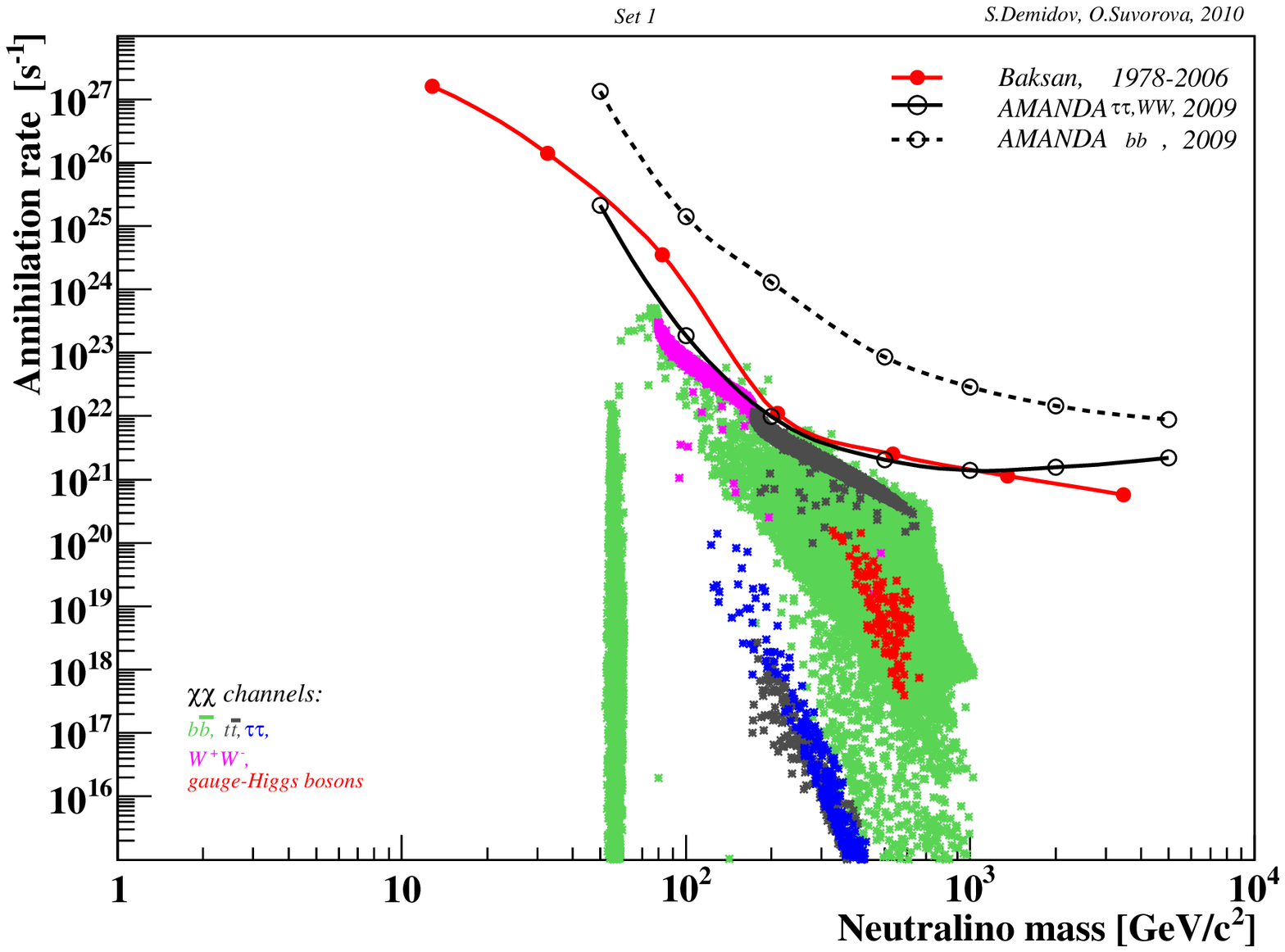}
\includegraphics[width=0.9\columnwidth,height=0.6\columnwidth]{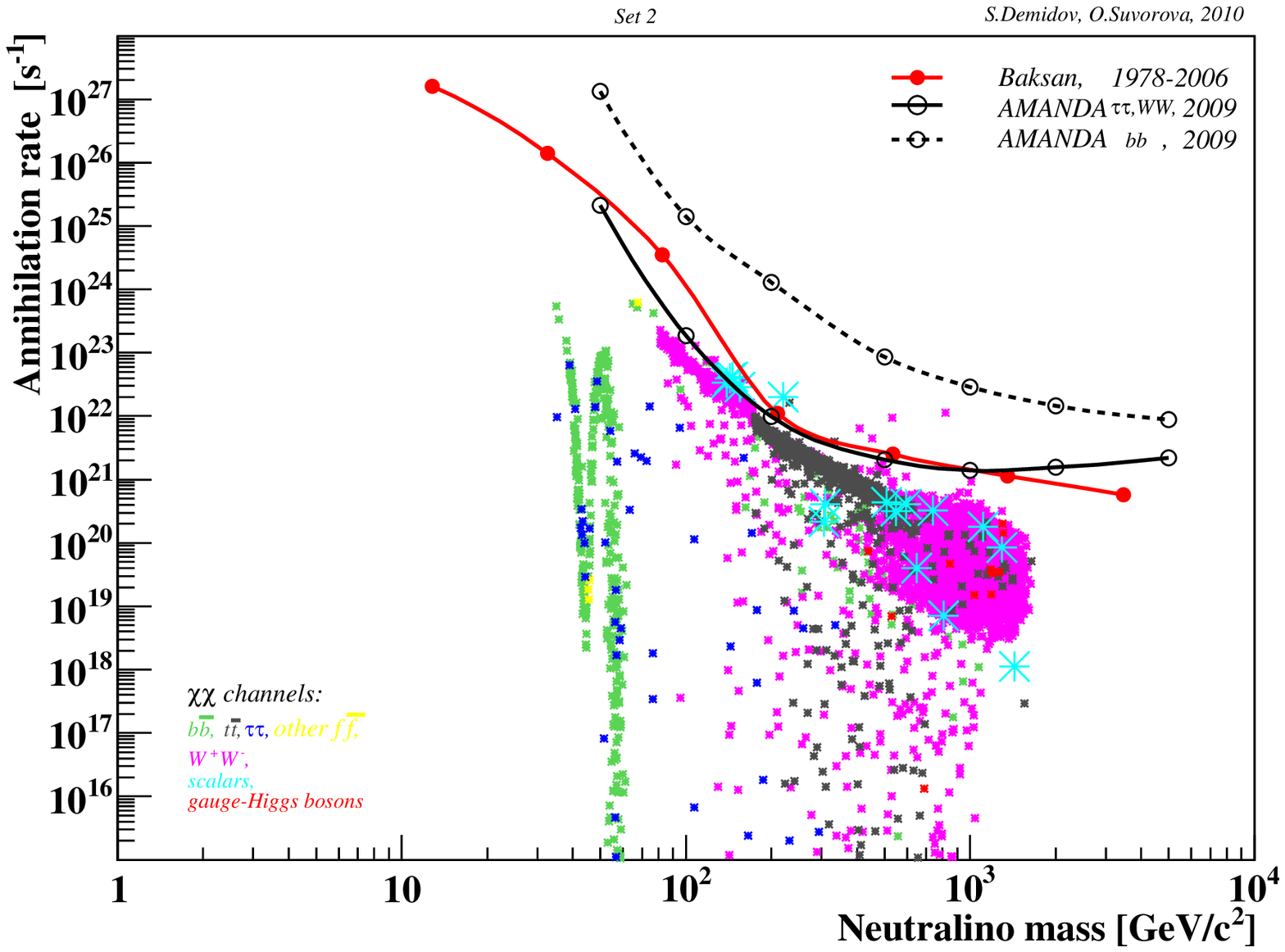}
\caption{\label{arate_wmap_annih}
Annihilation rates vs neutralino mass for {\it Set 1} (up) and {\it
  Set 2} (bottom) 
of parameters. Here all points are in $3\sigma$ WMAP interval:
$0.0913<\Omega_{DM} < 0.1285$. We show the dominant annihilation
channel (Br~$>$~0.5) by different colours.}
}
\FIGURE[htb]{
\includegraphics[width=0.9\columnwidth,height=0.6\columnwidth]{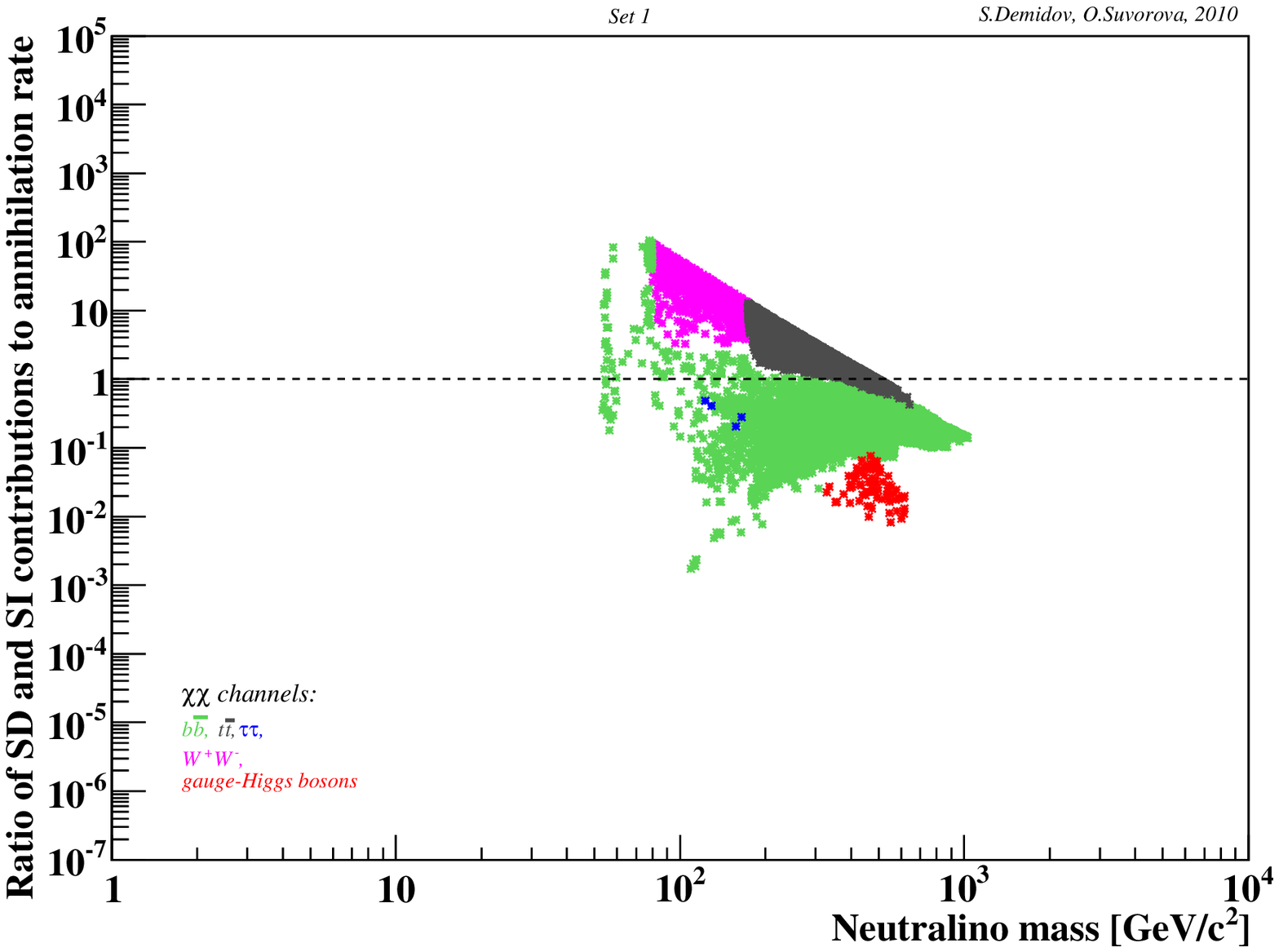}
\includegraphics[width=0.9\columnwidth,height=0.6\columnwidth]{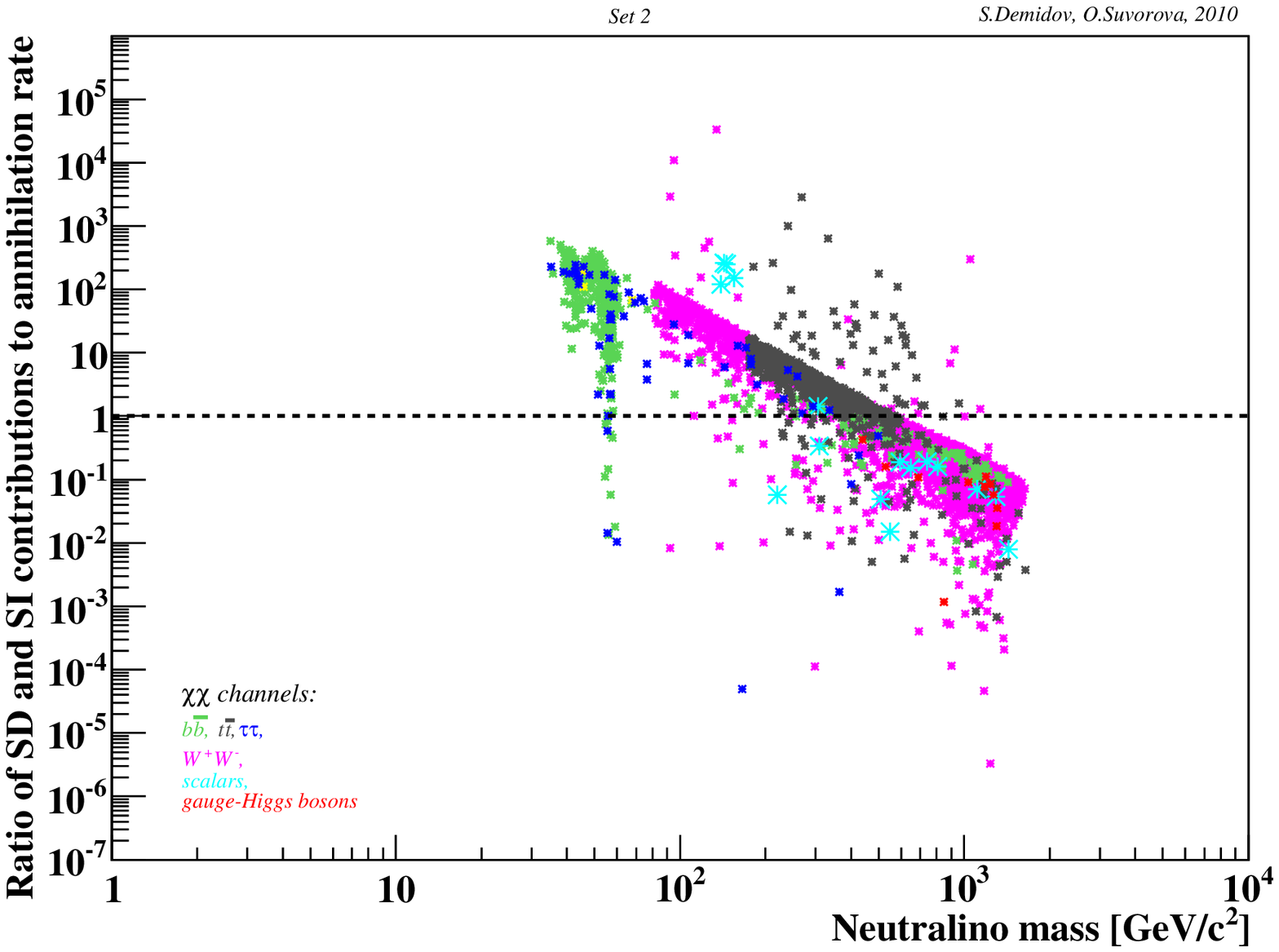}
\caption{\label{si_sd_compare}
Comparison of relative contribution of spin-dependent and
spin-independent interactions to annihilation rate. All conventions
are the same as in Fig.~\ref{arate_wmap_annih}.}
}
\FIGURE[htb]{
\includegraphics[width=0.9\columnwidth,height=0.6\columnwidth]{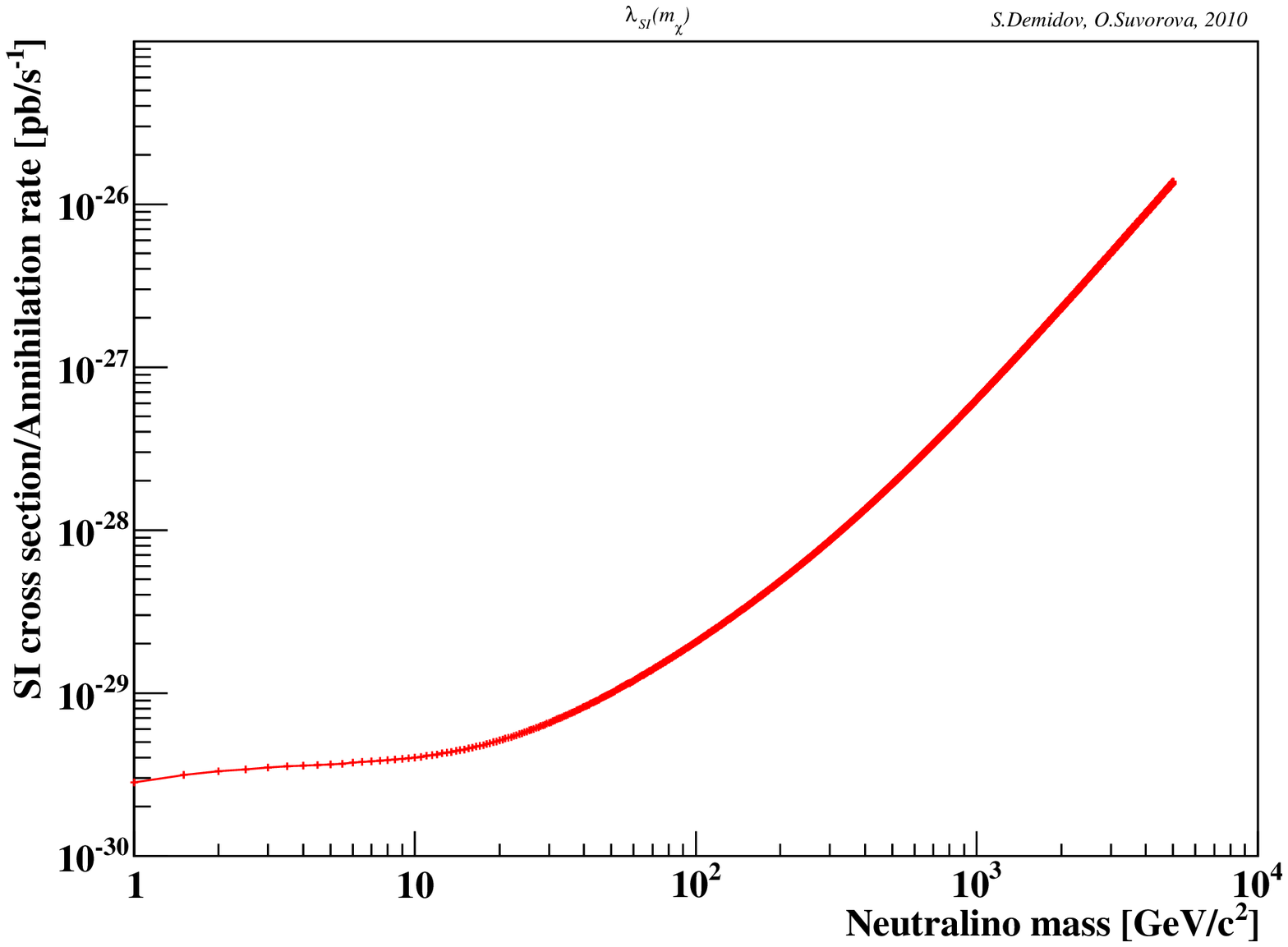}
\includegraphics[width=0.9\columnwidth,height=0.6\columnwidth]{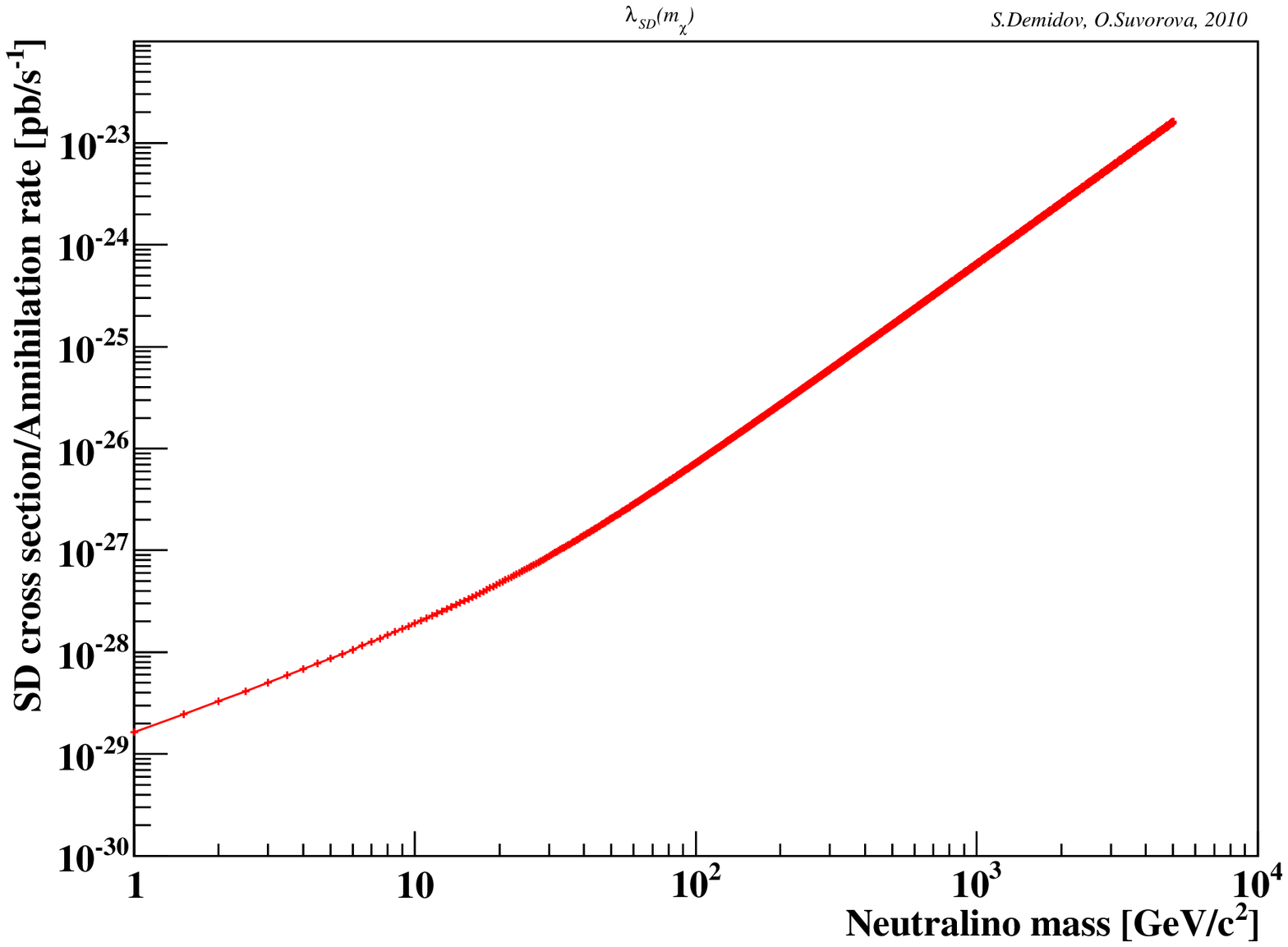}
\caption{\label{s_to_arate}
Coefficients $\lambda_{SI}(m_{\chi})$ (up) and
$\lambda_{SD}(m_{\chi})$ (bottom).
}
}
\FIGURE[htb]{
\includegraphics[width=0.9\columnwidth,height=0.6\columnwidth]{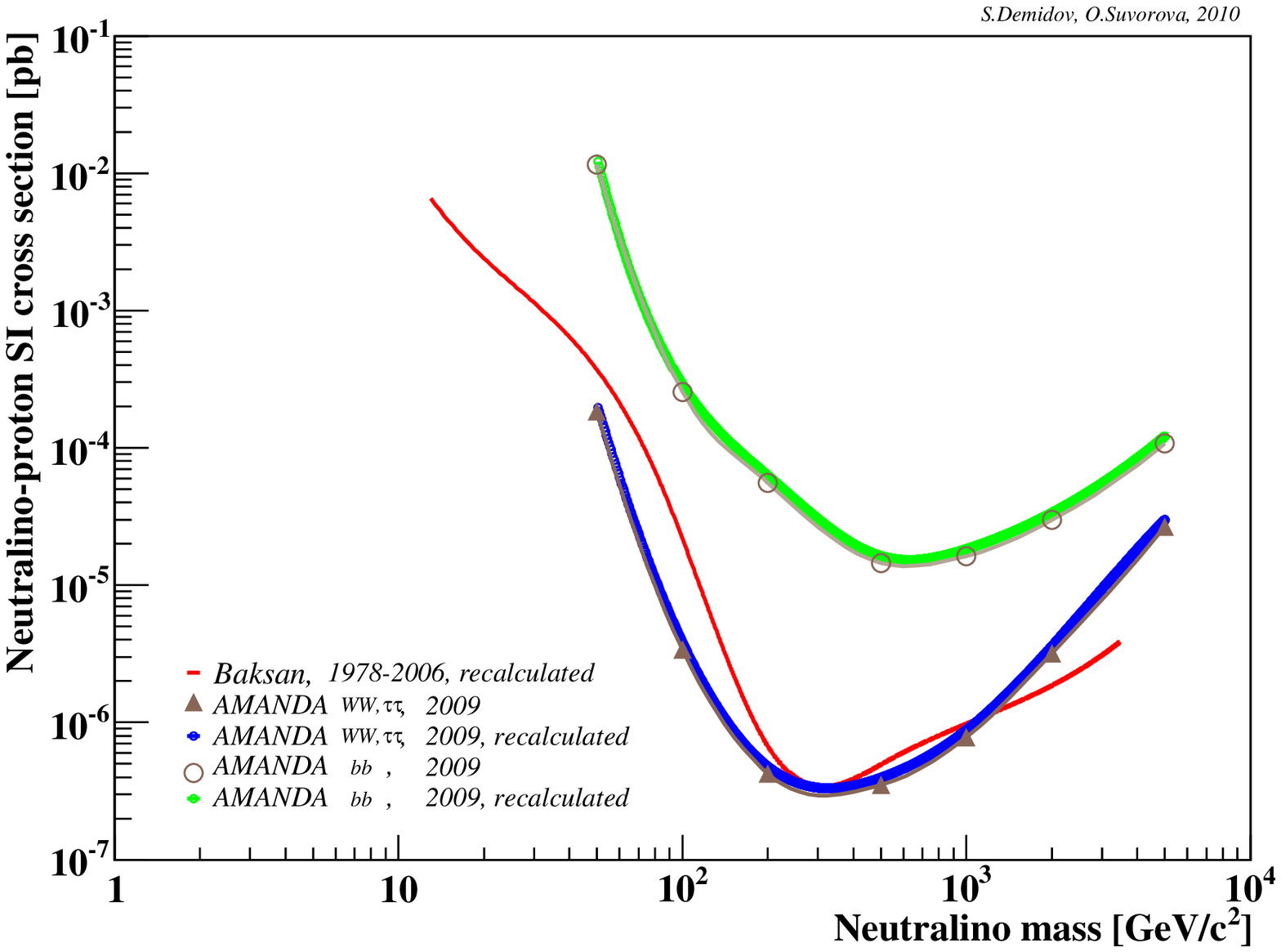}
\includegraphics[width=0.9\columnwidth,height=0.6\columnwidth]{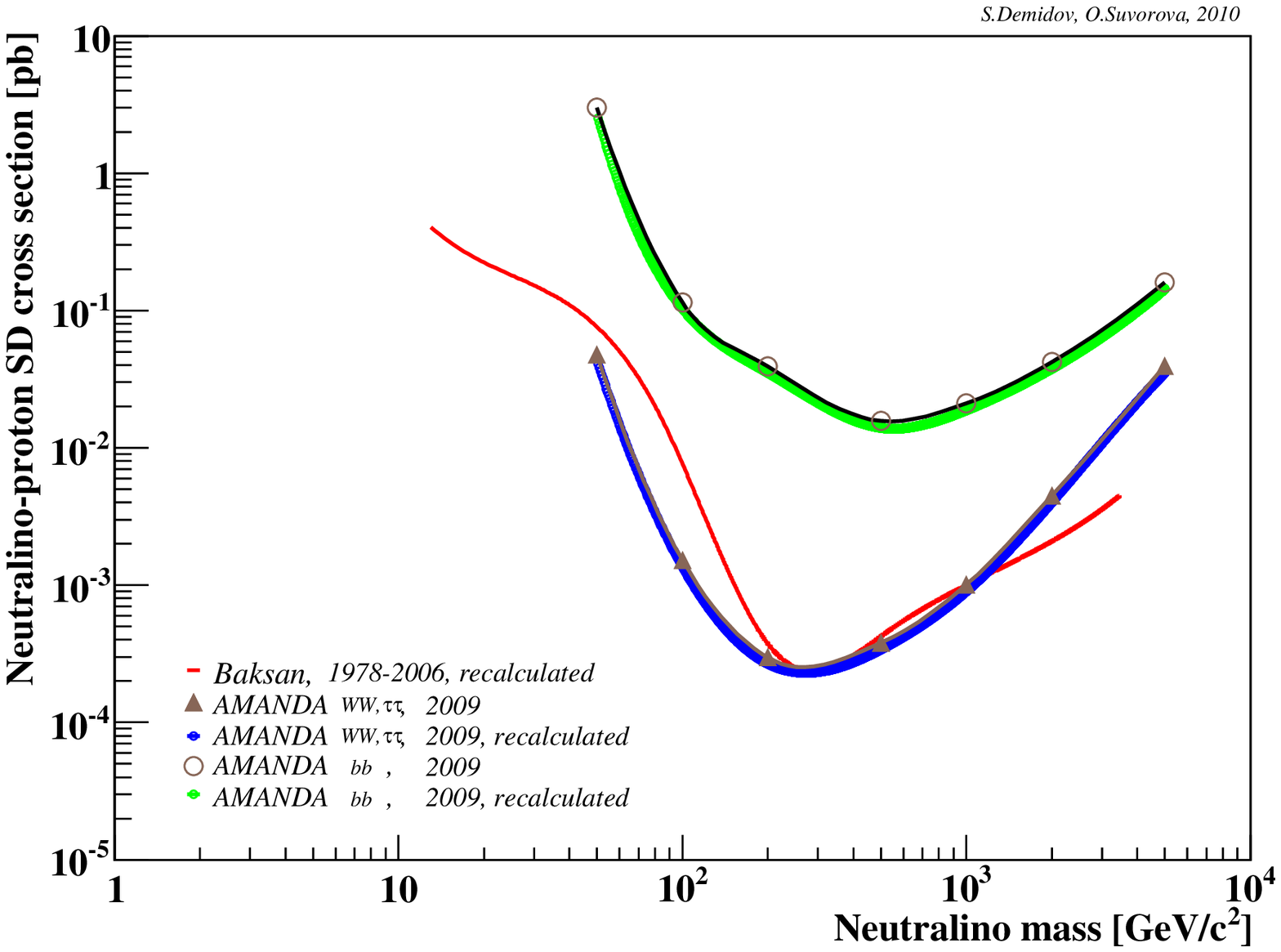}
\caption{\label{arate_koeff}
The upper limits at 90\% c.l. on neutralino-proton spin-independent (up) and
spin-dependent (bottom) cross sections for the Baksan underground
scintillator telescope recalculated from their 90\% c.l. limits on
annihilation rates~\cite{Baksan:06} (red lines). Also shown are the
AMANDA upper limits at 90\% c.l. on SI and SD cross sections presented
in~\cite{AMANDAII} (triangles) and our recalculated values  
from the AMANDA upper limits on annihilation rates (blue lines). 
}
}
\FIGURE[htb]{
\includegraphics[width=0.9\columnwidth,height=0.6\columnwidth]{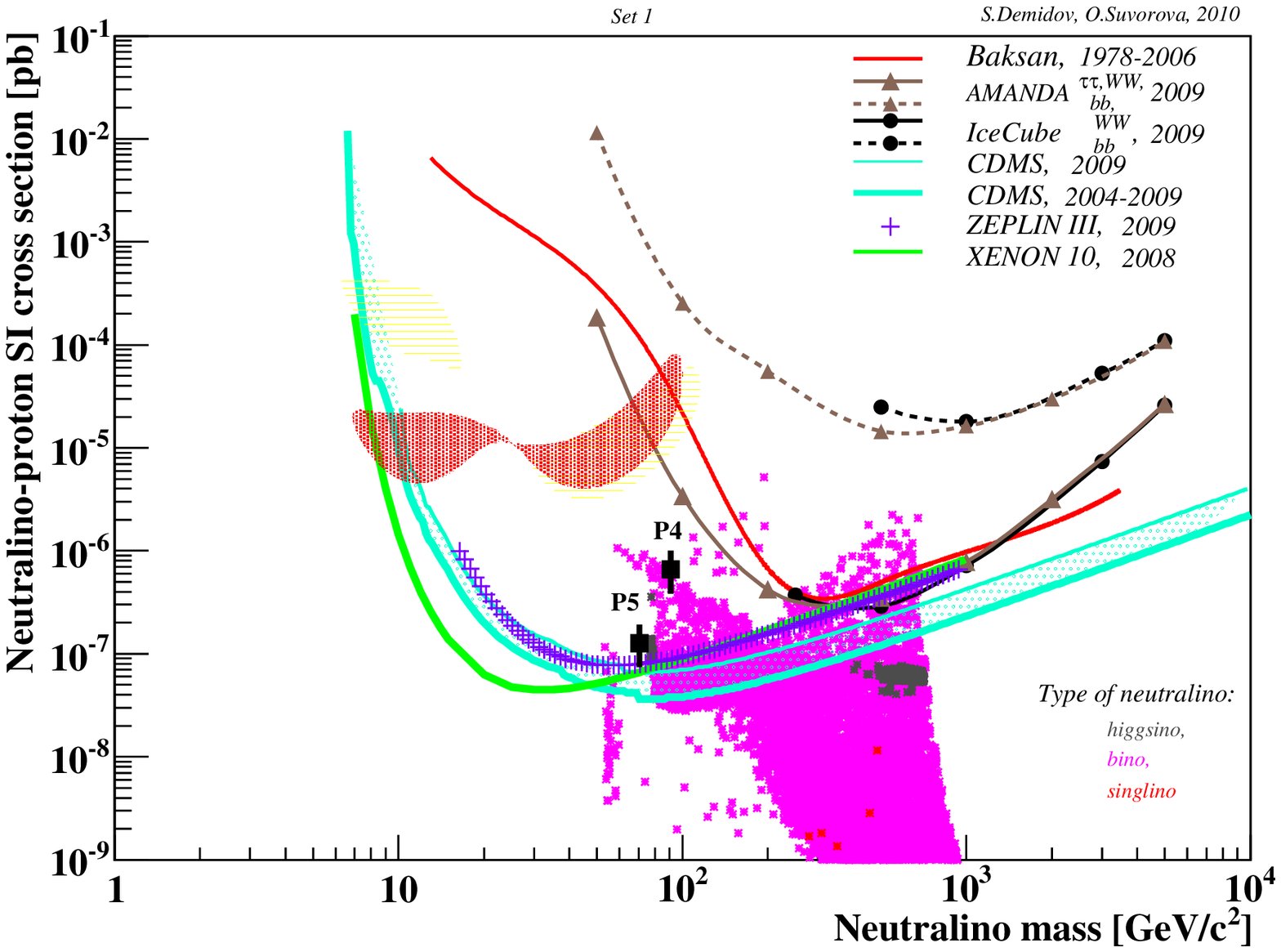}
\includegraphics[width=0.9\columnwidth,height=0.6\columnwidth]{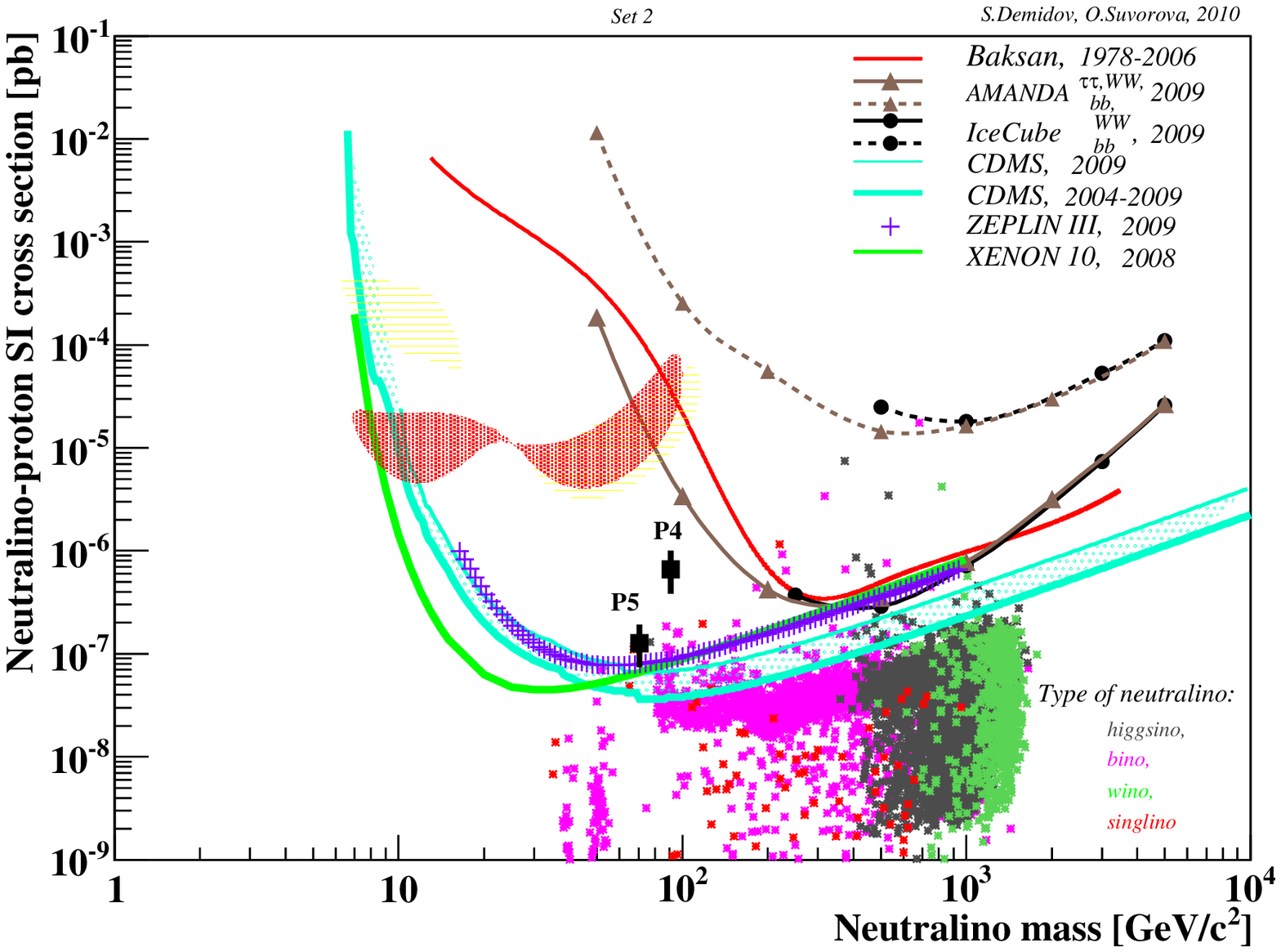}
\caption{\label{si_wmap}
Neutralino-proton spin-independent cross section vs neutralino mass for {\it Set 1} (up) and
{\it Set 2} (bottom) of parameters and experimental limits at 90\% c.l.
Neutralino content is shown by different colours.
}
}
\FIGURE[htb]{
\includegraphics[width=0.9\columnwidth,height=0.6\columnwidth]{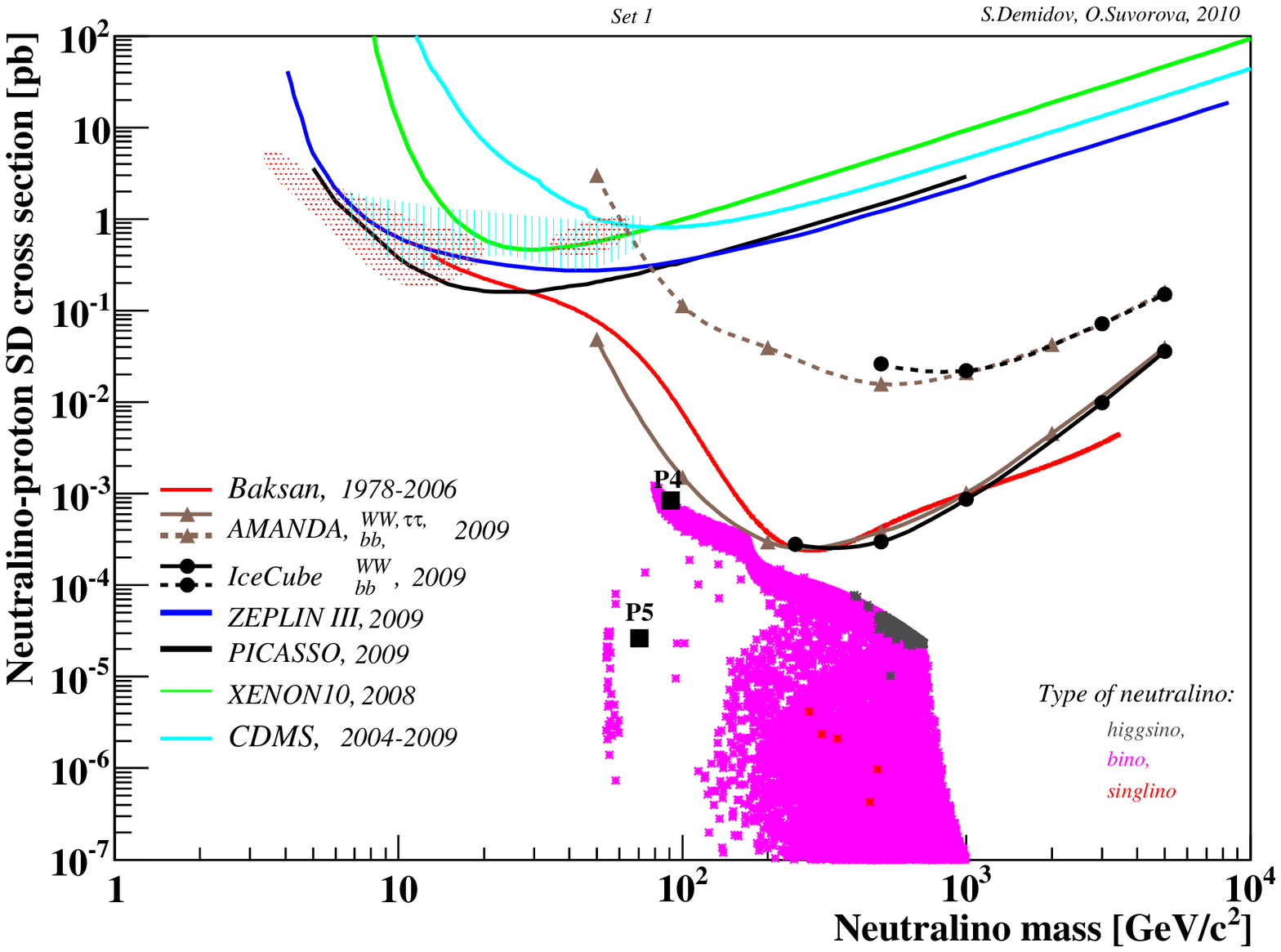}
\includegraphics[width=0.9\columnwidth,height=0.6\columnwidth]{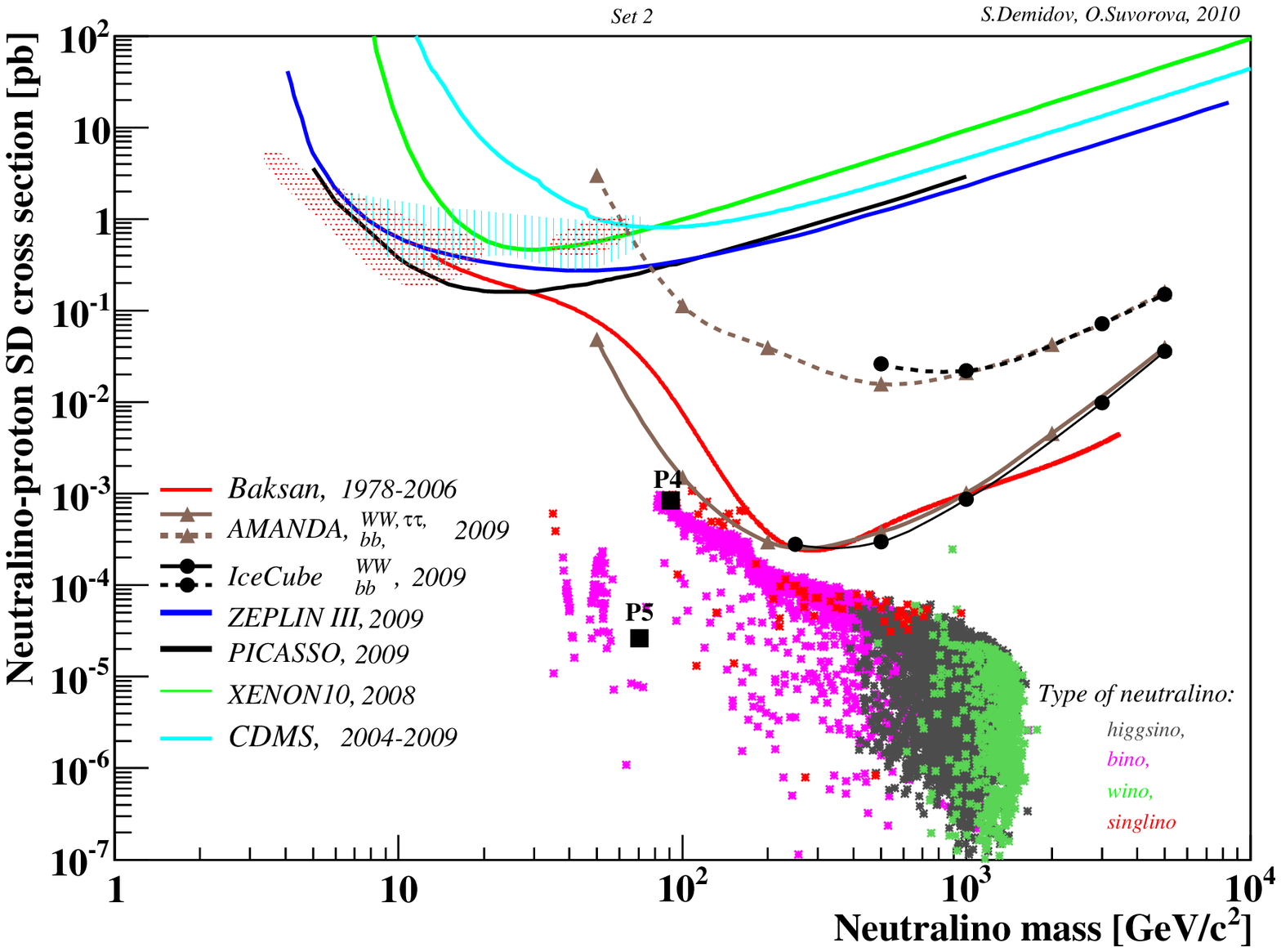}
\caption{\label{sd_wmap}
Neutralino-proton spin-dependent cross section vs neutralino mass for {\it Set 1} (up) and
{\it Set 2} (bottom) of parameters and experimental limits. The
notations are as in Fig.\ref{si_wmap}. All points are
constraint by SI cross section experimental limits at 90\% c.l.
}
}

%
%

\addtocontents{toc}{  }


\begin{thebibliography}{999}

\bibitem{WMAP:06} D.N.~Spergel {\em et al.} [WMAP Collab.], 
``Wilkinson Microwave Anisotropy Probe (WMAP) three year results:
  implications for cosmology,'' Astrophys.J.Suppl. {\bfseries 170}
  (2007) 377 [arXiv:astro-ph/0603449].

\bibitem{GravLin:98} J.A.~Tyson {\em et al.}, ``Detailed mass map of
  CL0024+1654 from strong lensing,'' Astrophys.~J. {\bfseries 498}
  (1998) L107 (1998) [arXiv:astro-ph/9801193]; H.~Dahle, ``A
  compilation of weak gravitational lensing studies of clusters of
  galaxies,'' [arXiv:astro-ph/0701598]. 

\bibitem{DClaster:01} A.~Borriello and P.Salucci {\em et al.}, ``The
  dark matter distribution in disc galaxies,''
  Mon.~Not.~Roy.~Astron.~Soc. {\bfseries 323} (2001) 285; M.~Persic
  {\em et al.}, ``The Universal rotation curve of spiral galaxies:
  1. The Dark matter connection,''
  Mon.~Not.~Roy.~Astron.~Soc. {\bfseries 281} (1996) 27.  

\bibitem{Zwicky:33} F.~Zwicky, ``Die Rotverschiebung von
  extragalaktischen Nebeln,'' Helv.~Acta 6 (1933) 110.

\bibitem{WMAP:09} E.~Komatsu {\em et al.} [WMAP Collab.], 
``Five-Year Wilkinson Microwave Anisotropy Probe (WMAP) Observations:
  Cosmological Interpretation,'' Astrophys.J.Suppl. {\bfseries 180},
  (2009) 330  [arXiv:astro-ph/0803.0547].

\bibitem{WIMP:85} S.~Steigman and M.~S.~Turner, ``Cosmological
  Constraints on the Properties of Weakly Interacting Massive
  Particles,'' Nucl.~Phys. {\bfseries B253} (1985) 375.

\bibitem{Jungman:96} G.~Jungman, M.~Kamionkowski and K.~Griest.
``Supersymmetric dark matter,'' Phys.\ Rept.\ {\bfseries 267} (1996)
  195 [arXiv:hep-ph/9506380].

\bibitem{Silk:05}
G.~Bertone, D.~Hooper, J.~Silk, ``Particle dark matter: Evidence,
candidates and constraints,'' Phys.\ Rept.\ {\bfseries 405} (2005) 279 
[arXiv:hep-ph/0404175].

\bibitem{DAMA:08} R.~Bernabei et al. [DAMA Collab.], ``First results
  from DAMA/LIBRA and the combined results with DAMA/NaI,''
  Eur.~Phys.~J. C {\bfseries 56} (2008) 333
  [arXiv:astro-ph/0804.2741]. 

\bibitem{CDMS:09} Z.~Ahmed et al. [CDMS Collab.], ``Search for Weakly
  Interacting Massive Particles with the First Five-Tower Data from
  the Cryogenic Dark Matter Search at the Soudan Underground
  Laboratory,'' Phys.~Rev.~Lett.{\bfseries 102} (2009) 011301; 
  Z.~Ahmed et al. [CDMS Collab.], ``Results from the Final Exposure of  
  the  CDMS II Experiment,'' [arXiv:astro-ph/0912.3592].

\bibitem{WMAP:08} G.~Dobler and D.~P.~Finkbeiner, ``Extended Anomalous
  Foreground Emission in the WMAP 3-Year Data,'' Astrophys.~J. {\bfseries 680} (2008) 1222 
 [arXiv:astro-ph/0712.1038].  

\bibitem{EGRET:06} W.~de Boer {\em et al.}, ``The supersymmetric
  interpretation of the EGRET excess of diffuse Galactic gamma rays,''
  Phys.~Lett. {\bfseries 636} (2006) 13 [arXiv:astro-ph/0511154].

\bibitem{FERMI:09} W.B.~Atwood {\em et al.} [Fermi/LAT Collab.], ``The
  Large Area Telescope on the Fermi Gamma-ray Space Telescope
  Mission,'' Ap.~J. {\bfseries 697} No 2 (2009) 1071; 
V.~Vitale {\em et al.} [Fermi/LAT Collab.], ``Indirect Search for Dark
  Matter from the center of the Milky Way with the Fermi-Large Area
  Telescope,'' Proc. of 2009 Fermi Symposium, Washington, D.C.,
  (2009) [arXiv:astro-ph/0912.3828]. 

\bibitem{INTEGRAL:03} P.~Jean {\em et al.} [SPI/INTEGRAL Collab.],
  ``Early SPI/INTEGRAL measurements of galactic 511 keV line emission
  from positron annihilation,'' Astron.~Astrophys. {\bfseries 407}
  (2003) L55 [arXiv:astro-ph/0309484]; N.~Prantzos, ``On the
  intensity and spatial morphology of the 511 keV emission in the
  Milky Way,'' Astron.~Astrophys. {\bfseries 449} (2006) 869
  [arXiv:astro-ph/0511190].

\bibitem{PAMELA:08} O.~Adriani {\em et al.} [PAMELA Collab.],
  ``Observation of an anomalous positron abundance in the cosmic
  radiation,'' Nature {\bfseries 458} (2009) 607 [arXiv:astro-ph/0810.4995].  

\bibitem{ATIC:08} J.~Chang {\em et al.} [ATIC Collab.], ``An excess of
  cosmic ray electrons at energies of 300-800 GeV,'' Nature {\bfseries
  456} (2008) 362. 

\bibitem{Bottino:2008mf}
  A.~Bottino, F.~Donato, N.~Fornengo and S.~Scopel,
  ``Interpreting the recent results on direct search for dark matter particles
  in terms of relic neutralino,''
  Phys.\ Rev.\  D {\bf 78} (2008) 083520
  [arXiv:0806.4099 [hep-ph]].

\bibitem{Bottino:09}
V. Niro, A. Bottino, N. Fornengo, S. Scopel,
``Investigating light neutralinos at neutrino telescopes,''
 Phys.\ Rev. {\bf D80} (2009) 095019 [arXiv:hep-ph/09092348].

\bibitem{nmssmtools}
http://www.th.u-psud.fr/NMHDECAY/nmssmtools.html

\bibitem{Ellwanger:2005dv}
  U.~Ellwanger and C.~Hugonie,
  ``NMHDECAY 2.0: An Updated program for sparticle masses, Higgs masses,
  couplings and decay widths in the NMSSM,''
  Comput.\ Phys.\ Commun.\  {\bf 175} (2006) 290
  [arXiv:hep-ph/0508022].

\bibitem{Ellwanger:2006rn}
  U.~Ellwanger and C.~Hugonie,
  ``NMSPEC: A Fortran code for the sparticle and Higgs masses in the NMSSM with
  GUT scale boundary conditions,''
  Comput.\ Phys.\ Commun.\  {\bf 177} (2007) 399
  [arXiv:hep-ph/0612134].

\bibitem{Djouadi:2008uw}
  A.~Djouadi {\it et al.},  ``Benchmark scenarios for the NMSSM,''
   JHEP {\bf 0807} (2008) 002  [arXiv:0801.4321 [hep-ph]].

\bibitem{Edsjo:09} 
  G.~Wikstrom and J.~Edsjo,
  ``Limits on the WIMP-nucleon scattering cross-section from neutrino
  telescopes,'' JCAP {\bf 0904} (2009) 009
  [arXiv:0903.2986 [astro-ph.CO]].

\bibitem{Hugonie:2007vd}
  C.~Hugonie, G.~Belanger and A.~Pukhov, ``Dark Matter in the
  Constrained NMSSM,'' JCAP {\bf 0711} (2007) 009 [arXiv:0707.0628
  [hep-ph]]. 

\bibitem{Belanger:2008nt}
  G.~Belanger, C.~Hugonie and A.~Pukhov,
  ``Precision measurements, dark matter direct detection and LHC Higgs
  searches in a constrained NMSSM,'' JCAP {\bf 0901} (2009) 023
  [arXiv:0811.3224 [hep-ph]]. 

\bibitem{Ferrer:2006hy}
  F.~Ferrer, L.~M.~Krauss and S.~Profumo,
  ``Indirect detection of light neutralino dark matter in the NMSSM,''
  Phys.\ Rev.\  D {\bf 74} (2006) 115007 [arXiv:hep-ph/0609257].

\bibitem{Asplund:2009fu}
  M.~Asplund, N.~Grevesse, A.~J.~Sauval and P.~Scott, ``The chemical
  composition of the Sun,'' Ann.\ Rev.\ Astron.\ Astrophys.\  {\bf
  47} (2009) 481 [arXiv:0909.0948 [astro-ph.SR]].

\bibitem{Belanger:2008sj}
  G.~Belanger, F.~Boudjema, A.~Pukhov and A.~Semenov, ``Dark matter
  direct detection rate in a generic model with micrOMEGAs2.1,'' 
  Comput.\ Phys.\ Commun.\  {\bf 180} (2009) 747 [arXiv:0803.2360 [hep-ph]].  

\bibitem{sigma_pi} M.~M.~Pavan, I.~I.~Strakovsky, R.~L.~Workman,
  R.~A.~Arndt, ``The pion-nucleon Sigma term is definitely large:
  results from a G.W.U. analysis of pion nucleon scattering data,''
  PiN Newslett. 16 (2002) 110-115 [arXiv:hep-ph/0111066v1]

\bibitem{Gould:1987ir}
  A.~Gould, ``Resonant Enhancements In Wimp Capture By The Earth,''
  Astrophys.\ J.\  {\bf 321} (1987) 571.

\bibitem{Edsjo:2010sun} S.~Sivertsson and J.~Edsjo,
``Accurate calculations of the WIMP halo around the Sun and prospects
  for its gamma-ray detection,'' Phys. Rev. D81 (2010) 063502 
[arXiv:astro-ph.SR/0910.0017].

\bibitem{Griest:1986yu}
  K.~Griest and D.~Seckel, ``Cosmic Asymmetry, Neutrinos and the
  Sun,'' Nucl.\ Phys.\  B {\bf 283} (1987) 681
  [Erratum-ibid.\  B {\bf 296} (1988) 1034].

\bibitem{Gould:1992}
  A.~Gould, ``Cosmological density of WIMPs from solar and terrestrial
  annihilations,''  Astrophys.\ J.\  {\bf 388} (1992) 338. 

\bibitem{Ellis:2008hf}
  J.~R.~Ellis, K.~A.~Olive and C.~Savage,
  ``Hadronic Uncertainties in the Elastic Scattering of Supersymmetric Dark
  Matter,''
  Phys.\ Rev.\  D {\bf 77} (2008) 065026
  [arXiv:0801.3656 [hep-ph]].

\bibitem{Ellis:2009ka}
  J.~Ellis, K.~A.~Olive, C.~Savage and V.~C.~Spanos,
  ``Neutrino Fluxes from CMSSM LSP Annihilations in the Sun,''
  [arXiv:0912.3137 [hep-ph]].

\bibitem{Nezri:2010}
  F.~S.~Ling, E.~Nezri, E.~Athanassoula and R.~Teyssier,
  ``Dark Matter Direct Detection Signals inferred from a Cosmological N-body
  Simulation with Baryons,''
  JCAP {\bf 1002} (2010) 012
  [arXiv:0909.2028 [astro-ph.GA]].

\bibitem{Baksan:06} M.M.Boliev {\em et al.}, ``Results with the Baksan
  neutrino telescope,'' Proc. of the First Workshop on Exotic Physics
  with Neutrino Telescope, Uppsala, Sweden, 19 (2006)
  [arXiv:astro-ph/0701333].

\bibitem{AMANDAII}
  J.~Braun and D.~Hubert for the IceCube~Collaboration,
  ``Searches for WIMP Dark Matter from the Sun with AMANDA,''
  arXiv:0906.1615 [astro-ph.HE].

\bibitem{Baksan:96} M.M.Boliev et al., ``Search for supersymmetric
  dark matter with Baksan Underground Telescope,'' Nucl. Phys.,
  {\bfseries 48} (1996) 83. 

\bibitem{Baksan:97} M.M.Boliev et al., ``BAKSAN NEUTRALINO SEARCH,''
  Dark Matter in Astro- and Particle Physics, edited by
  H.V.Klapdor-Kleingrothaus and Y.Ramachers (Singapore: World Sci.),
  (1997), 711; see also O.V.Suvorova, ``Status and Perspectives of
  Indirect Search for Dark Matter,'' [arXiv:hep-ph/9911415].

\bibitem{SuperK:04} S.~Desai {\em et al.}, ``Search for Dark Matter
  WIMPs using Upward Through-going Muons in Super-Kamiokande,'' Phys.\
  Rev.\ {\bfseries D70} (2004) 083523. 

\bibitem{IceCube} R. Abbasi {\em et al.} [Ice Cube Collab.], ``Limits
  on a muon flux from neutralino annihilations in the Sun with the
  IceCube 22-string detector,''Phys.Rev.Lett.{\bfseries 102} (2009)
  201302, see also http://www.icecube.wisc.edu/news/.

\bibitem{Angle:2007uj}
  J.~Angle {\it et al.}  [XENON Collaboration],
  ``First Results from the XENON10 Dark Matter Experiment at the Gran Sasso
  National Laboratory,''
  Phys.\ Rev.\ Lett.\  {\bf 100} (2008) 021303
  [arXiv:0706.0039 [astro-ph]].

\bibitem{Lebedenko:2008gb}
  V.~N.~Lebedenko {\it et al.},
  ``Result from the First Science Run of the ZEPLIN-III Dark Matter Search
  Experiment,''
  Phys.\ Rev.\  D {\bf 80} (2009) 052010
  [arXiv:0812.1150 [astro-ph]].

\bibitem{Archambault:2009}
  S.~Archambault {\it et al.},
  ``Dark Matter Spin-Dependent Limits for WIMP Interactions on 19-F by PICASSO,''
  Phys.\ Lett.\ B {\bf 682} (2009) 185  
[arXiv:0907.0307 [hep-ex]].


\bibitem{Baikal:09}
A.~Avrorin {\it et al.},
  ``Search for Neutrinos from Dark Matter Annihilation in the Sun with the 
Baikal Neutrino Experiment,''
Proc. of the 31st ICRC, Lodz, Poland (2009) [arXiv:0909.5589v1 [astro-ph.HE]]

\bibitem{ANTARES:09}
G.~Lim for the ANTARES Collaboration,
  ``First results on the search for dark matter in the Sun with the ANTARES neutrino telescope,''
Proc. of the 31st ICRC, Lodz, Poland (2009) [arXiv:0905.2316v3 [astro-ph.CO]]

\bibitem{DMTools:09}
http://dmtools.brown.edu

\bibitem{Cirelli:2005gh}
  M.~Cirelli, N.~Fornengo, T.~Montaruli, I.~Sokalski, A.~Strumia and F.~Vissani,
  ``Spectra of neutrinos from dark matter annihilations,''
  Nucl.\ Phys.\  B {\bf 727} (2005) 99
  [Erratum-ibid.\  B {\bf 790} (2008) 338]
  [arXiv:hep-ph/0506298];
  M.~Blennow, J.~Edsjo and T.~Ohlsson,
  ``Neutrinos from WIMP Annihilations Using a Full Three-Flavor Monte Carlo,''
  JCAP {\bf 0801} (2008) 021
  [arXiv:0709.3898 [hep-ph]].


\end{thebibliography}
\end{document}